\documentclass[review]{elsarticle}

\usepackage{lineno,hyperref}
\usepackage{amsmath}
\usepackage{subcaption} 
\usepackage{IEEEtrantools}

\usepackage[usenames,dvipsnames]{xcolor} 

\modulolinenumbers[5]

\journal{Journal of Computational Physics}

\newcommand\headingtable{%
\begin{tabular}[b]{l}
	Preprint submitted to Journal of Computational Physics\hfill\today\\\\
	\copyright2017. This manuscript version is made available under the CC-BY-NC-ND 4.0 license\\
	\url{http://creativecommons.org/licenses/by-nc-nd/4.0/}
\end{tabular}}
\makeatletter
\def\ps@pprintTitle{%
     \let\@oddhead\@empty
     \let\@evenhead\@empty
     \def\@oddfoot{\footnotesize\itshape\headingtable}%
     \let\@evenfoot\@oddfoot}
\makeatother

\newcommand{\mbf}[1]{\mathbf{#1}}
\newcommand{\mbs}[1]{\boldsymbol{#1}}
\newcommand{\mrm}[1]{\mathrm{#1}}

\biboptions{authoryear}

\begin{document}

\begin{frontmatter}

\title{A collision model for grain-resolving simulations of flows over dense, mobile, polydisperse granular sediment beds}

\author{Edward Biegert \fnref{edfootnote,mymainaddress}}
\author{Bernhard Vowinckel \fnref{bernfootnote,mymainaddress}}
\author{Eckart Meiburg \fnref{eckfootnote,mymainaddress}}
\address[mymainaddress]{Department of Mechanical Engineering, University of California, Santa Barbara, CA, USA}
\fntext[edfootnote]{ebiegert@engineering.ucsb.edu}
\fntext[bernfootnote]{vowinckel@engineering.ucsb.edu}
\fntext[eckfootnote]{meiburg@engineering.ucsb.edu}

%
%

\begin{abstract}
We present a collision model for phase-resolved Direct Numerical Simulations of sediment transport that couple the fluid and particles by the Immersed Boundary Method.  Typically, a contact model for these types of simulations comprises a lubrication force for particles in close proximity to another solid object, a normal contact force to prevent particles from overlapping, and a tangential contact force to account for friction.  Our model extends the work of previous authors to improve upon the time integration scheme to obtain consistent results for particle-wall collisions. Furthermore, we account for polydisperse spherical particles and introduce new criteria to account for enduring contact, which occurs in many sediment transport situations.  This is done without using arbitrary values for physically-defined parameters and by maintaining the full momentum balance of a particle in enduring contact. We validate our model against several test cases for binary particle-wall collisions as well as the collective motion of a sediment bed sheared by a viscous flow, yielding satisfactory agreement with experimental data by various authors.
\end{abstract}


\end{frontmatter}

\section{Introduction}\label{ref:introduction}

The flow over dense, mobile granular beds plays a central role in multiple applications in environmental, mechanical, and process engineering.
Prime examples of this type of problem are turbidity currents and powder snow avalanches \citep{meiburg2010}, for which resuspension of particles essentially determines the dynamics of the flow. The resuspension threshold is quantified by the ratio of hydrodynamic drag and lift forces to the weight of the particles, known as the Shields number \citep{shields1936}. The critical Shields parameter, however, has proven to be a poor predictor for the onset of particle erosion \citep{garcia2008}, and substantial efforts have been made in hydraulic engineering to overcome this difficulty \citep{seminara2010}. To date, progress has been slow due to the experimental difficulty of measuring dense particle-laden flows in a laboratory. In recent years, numerical simulations have provided an alternative way to study fluid-particle interactions under these complex conditions. In particular, the Direct Numerical Simulation (DNS) of particle-laden flows using the Immersed Boundary Method (IBM) has become a very attractive option. A main advantage of this approach is that it allows for a fully-coupled system by accounting for particle-fluid, fluid-particle, and particle-particle interactions. While coupling between the particle and the fluid can be realized by a straightforward implementation of an IBM, particle-particle coupling has to be modeled by suitable expressions for the collision processes involved.

A number of different phase-resolving methods to simulate mobile particles in a viscous flow have been developed in the past two decades.  \cite{glowinski1999} and \cite{patankar2000} developed a Distributed Lagrange Multiplier (DLM)/fictitious domain approach that forces the fluid throughout the volume of the particle to move in a rigid body motion with the particle.  While this method was designed for a finite element framework, \cite{kajishima2001} and \cite{apte2009} later developed different versions to be used in a finite difference framework.  More recently, other methods that enforce the no-slip condition on the particle surface have been developed. \cite{zhang2005} developed PHYSALIS, which uses the analytical solution for Stokes flow around a sphere.  \cite{uhlmann2005} developed an IBM that enforces the no-slip condition using interpolation and spreading operations via Dirac delta functions.  \cite{kempe2012} extended this method to make it stable for a larger range of particle/fluid density ratios. Although there exists a variety of phase-resolving methods, the choice of collision models, on the other hand, has not been as diverse.
\cite{derksen2011} and \cite{derksen2015} used a hard-sphere model, which resolves collisions instantaneously.
\cite{glowinski1999} have developed a repulsive potential (RP) model that prevents particles from overlapping by applying a repulsive force at some small distance before the particles come in contact.  Many other authors have adopted this model for simulations involving dilute suspensions of particles.  For example, \cite{uhlmann2008} and \cite{santarelli2015} investigated particles in a vertical turbulent channel flow, \cite{lucci2010} studied the impact of finite size particles on isotropic turbulence, and \cite{breugem2012} and \cite{picano2015} have presented results for a horizontal flow laden with neutrally-buoyant particles.  In these simulations, particles rarely came in contact, and thus were successfully governed mostly by the IBM, using the repulsive potential only to prevent overlap.

For the situation involving shear flow over a densely-packed sediment bed, however, particle-particle contact becomes ubiquitous \citep{balachandar2010}.
Hard sphere models cannot maintain simultaneous collisions or enduring contacts between multiple interacting particles, but instead represent sediment beds as particles in constant, minute motion \citep{derksen2015}.  However, they have been used to reproduce critical erosion conditions for a laminar shear flow \citep{derksen2011}.
The drawbacks of the RP model for the situation of sediment transport have been clearly-elaborated by \cite{kempe2014}. Using the RP in the framework of the IBM introduces an artificial gap of two times the grid cell size between colliding particles so that the fluid in the gap between the particle surfaces can still be resolved \citep{glowinski2001}. For the situation of sediment transport, however, the artificial gap also introduces an unphysical protrusion of the particles into the horizontal flow, which is critical as the protrusion has been acknowledged to be a very sensitive parameter for particle mobilization \citep{fenton1977}. In addition, the RP model introduces a material stiffness $k_n$ which has to be calculated {\em a priori} to design a collision model that is numerically stable. If the value of $k_n$ is chosen too high, the repulsive force is overestimated and the particle would experience an unphysical high rebound velocity. On the other hand, if $k_n$ is too low, the duration of the particle collision would be too large. The resuspension mechanisms, however, generate a high variety of particle impact velocities $u_{in}$ for the collisions, typically characterized by the nondimensional Stokes number $\mrm{St}=\rho_p u_{in} D_p / (9\, \rho_f\, \nu_f)$, where $\rho_p$ and $\rho_f$ are the particle and fluid density, respectively, $D_p$ is the particle diameter, and $\nu_f$ is the kinematic viscosity of the fluid. In fact, for bed-load transport in water the variety can span from $\mrm{St}\gg 10$ \citep[saltating particles,][]{nino1998} to $\mrm{St} = \mathcal{O}(10)$ \citep[rolling particles,][]{lajeunesse2010} to $\mrm{St}\ll 1$ (enduring contact within the sediment bed). Hence, selecting a stiffness that is stable for high-impact velocities results in excessively low stiffnesses within the bed, which acts as an unphysical dampening of the system.

Although there are studies of particle-laden horizontal flows such as \cite{shao2012} and \cite{kidanemariam2013} in which the model by \cite{glowinski1999} has been employed, these had to be limited to small volume fractions and conclusions about particle-particle interaction have not been possible.
\cite{kidanemariam2014} used an improved repulsive linear spring-dashpot model that solves the issues with calibrating $k_n$, but still relies on an artificial gap distance.  Thus, in order to obtain appropriate bulk sediment transport quantities, they calibrated the dry restitution coefficient $e_{dry} = -u_{out} / u_{in}$, where $u_{out}$ is the rebound velocity as soon as the collision process is finished, although this parameter can be set exactly as a material property. It describes the dissipation of kinetic energy due to the inelastic mechanics of the dry contact and is typically in the range of $0.8 \leq e_{dry} <1$ for silicate materials \cite[e.g.][]{joseph2001}. In the study of \cite{kidanemariam2014}, a rather unphysical value of $e_{dry}=0.3$ was used to match the bulk transport rates of glass spheres from the experiments of \cite{aussillous2013}.

More recently, a more consistent approach has been advocated in the literature for which the artificial gap size is no longer needed \citep{simeonov2012,kempe2012a,izard2014,costa2015,sierakowski2016}. This approach uses a lubrication force when the particles come in close contact ($0 < \zeta_n \leq 2h$, where $\zeta_n$ is the distance between the two surfaces), and a contact force when the surfaces come in contact and slightly overlap  ($\zeta_n \leq 0$).  The lubrication force, which is based on lubrication theory, models the fluid forces acting on the particle that cannot be resolved by the computational mesh.  The contact force models material deformations and friction through components that are, respectively, normal and tangent to the surface.
Since these models attempt to address the actual physics of the collision, they have had much success in reproducing the desired restitution coefficients over a range of Stokes numbers for the experiments by \cite{joseph2001}.  \cite{simeonov2012} performed a detailed analysis breaking down the individual effects from lubrication forces, contact forces, and hydrodynamic forces.  However, only \cite{kempe2012a} and \cite{costa2015} have demonstrated that their models were able to reproduce the trajectories of a particle-wall impact provided by the benchmark experiments of \cite{gondret2002}.  Both studies show that the good agreement with the rebound trajectories is made possible by slightly stretching the collision process in time: long enough to resolve the response of the fluid field to the particle kinematics but shorter than any relevant physical timescale in the flow. In addition, both modeling approaches use an adaptive procedure to obtain mathematically-rigorous solutions to the ordinary differential equations governing particle motion during contact.  Using the model of \cite{kempe2012a}, a breakthrough was achieved by \cite{vowinckel2014}, who successfully carried out numerical simulations of turbulent horizontal channel flow laden with tens of thousands of particles.

In the present work, we build on the model proposed by \cite{kempe2012a} to extend it to situations of very dense packing fractions.  For example, in order for their collision model to work as designed, \cite{kempe2012a} neglect the hydrodynamic forces acting on a particle while it is in contact with another object.  This was addressed implicilty by \cite{kempe2014}, even though it was not stated in their paper, by including the hydrodynamic forces in the equation of motion regardless of the type of collision (Kempe \& Fr\"{o}hlich, 2016, private communication).
Furthermore, their model for the tangential contact force, which is designed to exactly enforce zero slip between particles, does not converge to a steady-state configuration for enduring contact.
While this model worked well for simulations involving thin beds of particles at higher Reynolds numbers and Stokes numbers, we would like to extend it to work for thick beds of particles at a range of different Stokes numbers.
At this point, models from the Discrete Element Method (DEM) community, who simulate dry granular flows, seem to be more appropriate, as they introduce a ``memory" of the friction required to reach steady-state conditions \citep{Zhu2008}. \cite{costa2015} proposed a scheme with an enhanced treatment of lubrication forces, which can also be applied to smaller Stokes numbers as well as a variety of impact angles $\psi_{in}$. In this reference, however, they neither considered the situation of enduring contact nor conducted a validation on a larger scale addressing the collective effects of particle motion.

Finally, another aspect that has received far less attention so far is the situation of a sediment consisting of particles with varying particle diameters. Interestingly, all of the references cited so far deal with spherical monodisperse particles.  To the knowledge of the authors the only study considering horizontal channel flows laden with polydisperse sediment has been performed by \cite{fukuoka2014} using a front-tracking technique, but neither did the authors account for the feedback of the particles on the flow nor have they provided a validation for the experimental standard benchmark test cases such as particles settling in an ambient fluid or colliding with a wall. The absence of studies addressing polydisperse sediment with fully coupled IBM simulations is ever the more surprising, since its impact has been acknowledged as a key issue in the development and evolution of bedforms by segregation effects as reviewed by \cite{charru2013}.

As a consequence, the present work aims to resolve the problems mentioned above. Among the key challenges identified are i) deriving collision models for polydisperse sediment, ii) avoiding the introduction of an artificial gap between colliding particles, iii) adaptively-calibrating the particle stiffness to simulate a wide range of Stokes numbers in a consistent manner, iv) introducing suitable criteria to extend existing models towards the numerically-challenging situation of enduring contact for both normal and oblique collisions, and v) minimizing the number of tunable parameters within the model framework. We achieve our goals by presenting an implementation of collision models for polydisperse sediment. We use the adaptive procedure proposed by \cite{kempe2012a} for normal forces and the tangential model of \cite{thornton2013}, which stems from DEM. Furthermore, we extend both of these approaches for the situation of enduring contact. In particular, for enduring contact, we took care to retain all the governing terms of the momentum balance of a particle, i.e. hydrodynamic forces, buoyant weight, and collision forces. This measure turns out to be crucial when simulating flows over sediment beds, as the Shields parameter is based on the ratio of hydrodynamic to buoyant forces. The proposed enhancements allow us to reproduce several laboratory benchmark test cases for binary collisions. In addition, we present a detailed validation of our simulation results with wall-normal profiles of the fluid and particle velocities as well as bulk flow quantities using the experimental data of \cite{aussillous2013}.

 The paper is structured as follows. We briefly recall the numerical method, including the fluid solver, IBM, and the structure of the collision model, in Section \ref{sec:numerical_method}, followed by the mathematical description of the collision model employed in Section \ref{sec:collision_modeling}. We then present necessary enhancements to the collision model to deal with small Stokes numbers (Section \ref{sec:enhancements}) and to simulate dense granular packings with the gross of the particles in enduring contact (Section \ref{sec:enduring_contact}). Subsequently, the enhanced model is validated for the collective motion of polydisperse sediment sheared by a laminar flow in Section~\ref{sec:bulk_flow}.

\section{Particle motion and four-way coupling in the framework of the Immersed Boundary Method} \label{sec:numerical_method}
\subsection{Fluid solver}
%
For the present simulations, we solve the unsteady Navier-Stokes equations for an incompressible Newtonian fluid, given by

\begin{equation} \label{eq:navier_stokes}
\frac{\partial{\textbf{u}}}{\partial{t}}+\nabla\cdot(\textbf{u}\textbf{u}) = -\frac{1}{\rho_f}\:\nabla p + \nu_f \nabla^2 \textbf{u} + \textbf{f}_\textit{IBM} \hspace{0.5cm},
\end{equation}
and the continuity equation, given by
\begin{equation}\label{eq:continuity}
\nabla\cdot\textbf{u}=0 \qquad ,     \hspace{0.5cm}
\end{equation}
on a uniform rectangular grid with grid cell size $\Delta x = \Delta y = \Delta z = h$. Here, $\textbf{u}=(u,v,w)^{T}$ designates
the fluid velocity vector in Cartesian components, $p$ the pressure, $\nu_f$ the kinematic viscosity, $t$ the time,
and $\textbf{f}_\textit{IBM}$ an artificial volume force introduced by the IBM \citep{mittal2005}.
This volume force, which acts on the right-hand side of \eqref{eq:navier_stokes} in the vicinity of the inter-phase boundaries, connects the motion of the particles to the fluid phase.
The transfer of quantities, such as force and velocity, between Eulerian points and Lagrangian points, i.e. between fluid points of
the regular background grid and points on the surface of the particle, is performed by interpolation and spreading operations via a weighted
sum of regularized Dirac delta functions \citep{uhlmann2005}, of which we use the 3-point stencil function of \cite{roma1999}. The source term $\textbf{f}_\textit{IBM}$ is computed in such a way that the no-slip condition at the
particle surface is satisfied. Time advancement is achieved by a fractional step method, a third-order explicit low-storage three-step Runge-Kutta (RK) scheme
is employed for the convective terms, and a second-order semi-implicit Crank-Nicolson scheme is used for the viscous terms \citep{fadlun2000}, which are solved with the conjugate-gradient method.  Spatial derivatives are evaluated using second-order central-differencing. The pressure is treated with a direct solver based on Fast Fourier Transformations (FFT).  Our code can handle a variety of boundary conditions at the different walls, including no-slip, slip, periodic, and inflow/outflow.
%

\subsection{Computation of particle motion}\label{sec:eom}
Within the framework of the IBM, we calculate the motion of each individual spherical particle
by solving an ordinary differential equation for its translational velocity $\textbf{u}_p=(u_p,v_p,w_p)^{T}$
       %
\begin{linenomath*}
       \begin{equation}\label{eq:part_lin}
        m_p\: \frac{\text{d}\textbf{u}_p}{\text{d} t} = \underbrace{\oint_{\Gamma_p} \boldsymbol{\tau} \cdot \textbf{n}\: {\text{d}A}}_{=\textbf{F}_{h,p}} +
        \underbrace{V_p\:( \rho_p-\rho_f )\: \textbf{g}}_{=\textbf{F}_{g,p}} + \textbf{F}_{c,p} \qquad ,
       \end{equation}
\end{linenomath*}
        and its angular velocity $\boldsymbol{\omega}_p=(\omega_{p,x},\omega_{p,y},\omega_{p,z})^{T}$
\begin{linenomath*}
       \begin{equation}\label{eq:part_ang}
       I_p \:\frac{ \text{d}\boldsymbol{\omega}_p}{\text{d} t} = \underbrace{\oint_{\Gamma_p} \textbf{r}\times(\boldsymbol{\tau}\cdot\textbf{n})\:{\text{d}A}}_{=\textbf{T}_{h,p}} + \textbf{T}_{c,p} \hspace{0.5cm}.
      \end{equation}
\end{linenomath*}
Here, $m_p$ is the particle mass, $\Gamma_p$ the fluid-particle interface, $\boldsymbol{\tau}$ the hydrodynamic stress tensor, $\rho_p$ the particle density, $V_p$ the particle volume, $g$ the gravitational acceleration, $I_p=8\pi\rho_p R_p^{5}/15$ the moment of inertia, and $R_p$ the particle radius. Furthermore, the vector $\textbf{n}$ is the outward-pointing normal on the interface $\Gamma_p$, $\textbf{r} = \textbf{x} - \textbf{x}_p$ is the position vector of the surface point with respect to the center of mass $\textbf{x}_p$ of a particle, and $ \textbf{F}_{c,p}$ and $\textbf{T}_{c,p}$ are the force and torque due to particle collisions, respectively.  Furthermore, note the designation of the hydrodynamic force and torque as $\textbf{F}_{h,p}$ and $\textbf{T}_{h,p}$, respectively, as well as $\textbf{F}_{g,p}$ the force due to gravity, which will be used in the following for brevity.

We employ the approach of \cite{kempe2012} for evaluating the IBM forces and solving \eqref{eq:part_lin} and \eqref{eq:part_ang}.  We validated the fluid-particle coupling of the method against experimental data of a sphere settling in an unbounded quiescent fluid \citep{mordant2000} as well as towards a wall \citep{tencate2002}.
The particle was initially at rest for both setups and accelerated downwards due to gravity. For the unbounded case, the particle reached a constant terminal velocity $u_\infty$.  The Reynolds number $\mrm{Re}_p = u_\infty D_p/\nu_f$ based on this settling velocity is $\mrm{Re}_p = 41$. For the wall-bounded case, the particle reached a terminal velocity but then decelerated as it approached the wall.  For now, we only consider the particle motion before impact, which is governed by the IBM.  The collision model governing the impact will be described and validated further below in the text.  The Reynolds number based on the velocity before deceleration is $\mrm{Re}_p = 12$.  In both cases, the particle was discretized with 20 grid cells per diameter.
The respective data are plotted in Figure \ref{fig:settling_validation}, showing excellent agreement.
\setlength{\unitlength}{1cm}
\begin{figure}[t]
\begin{picture}(7,4.5)
  \put(0,  -0){\includegraphics[width=0.5\textwidth]{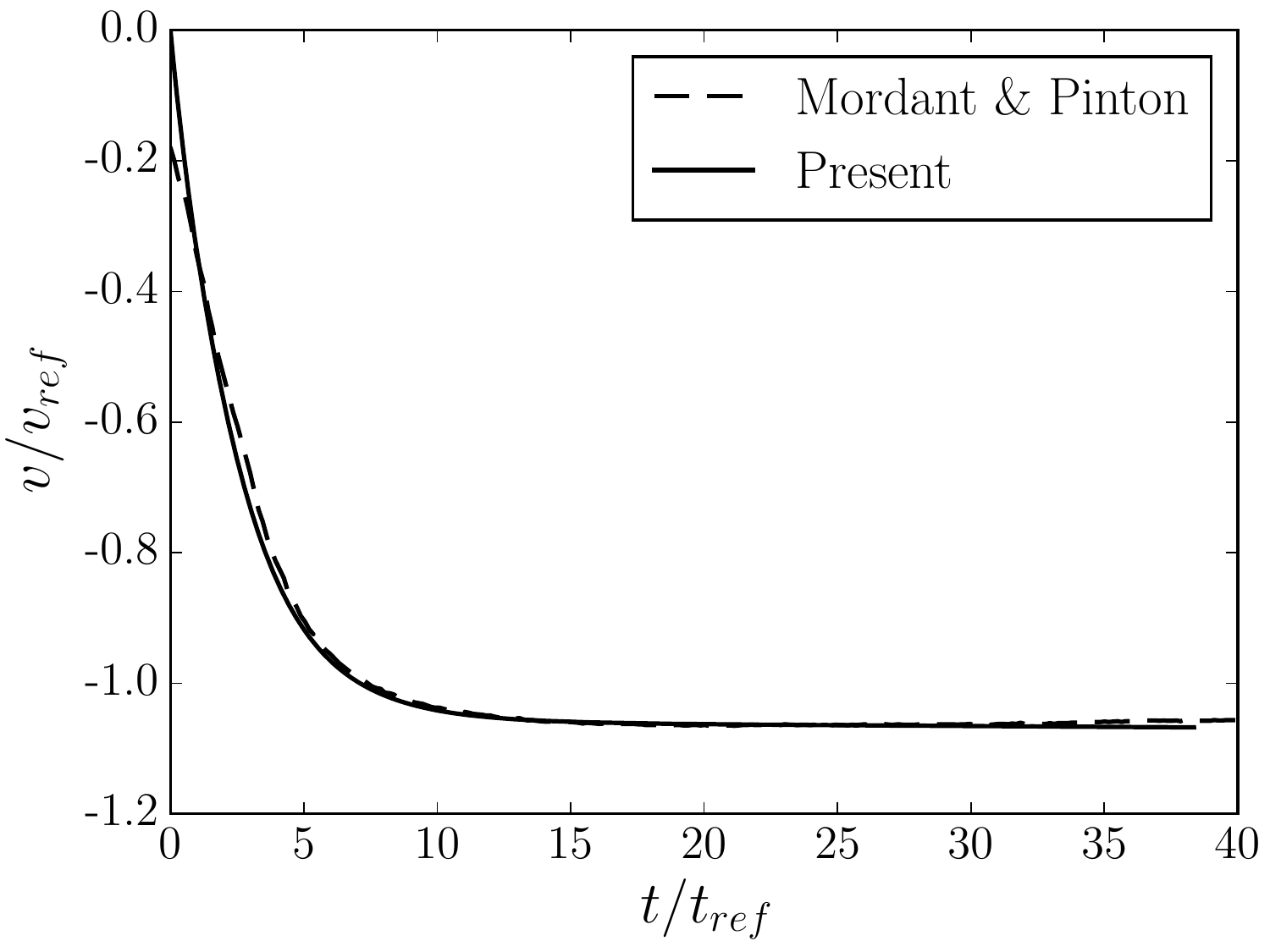}}
  \put(6.5,-0){\includegraphics[width=0.5\textwidth]{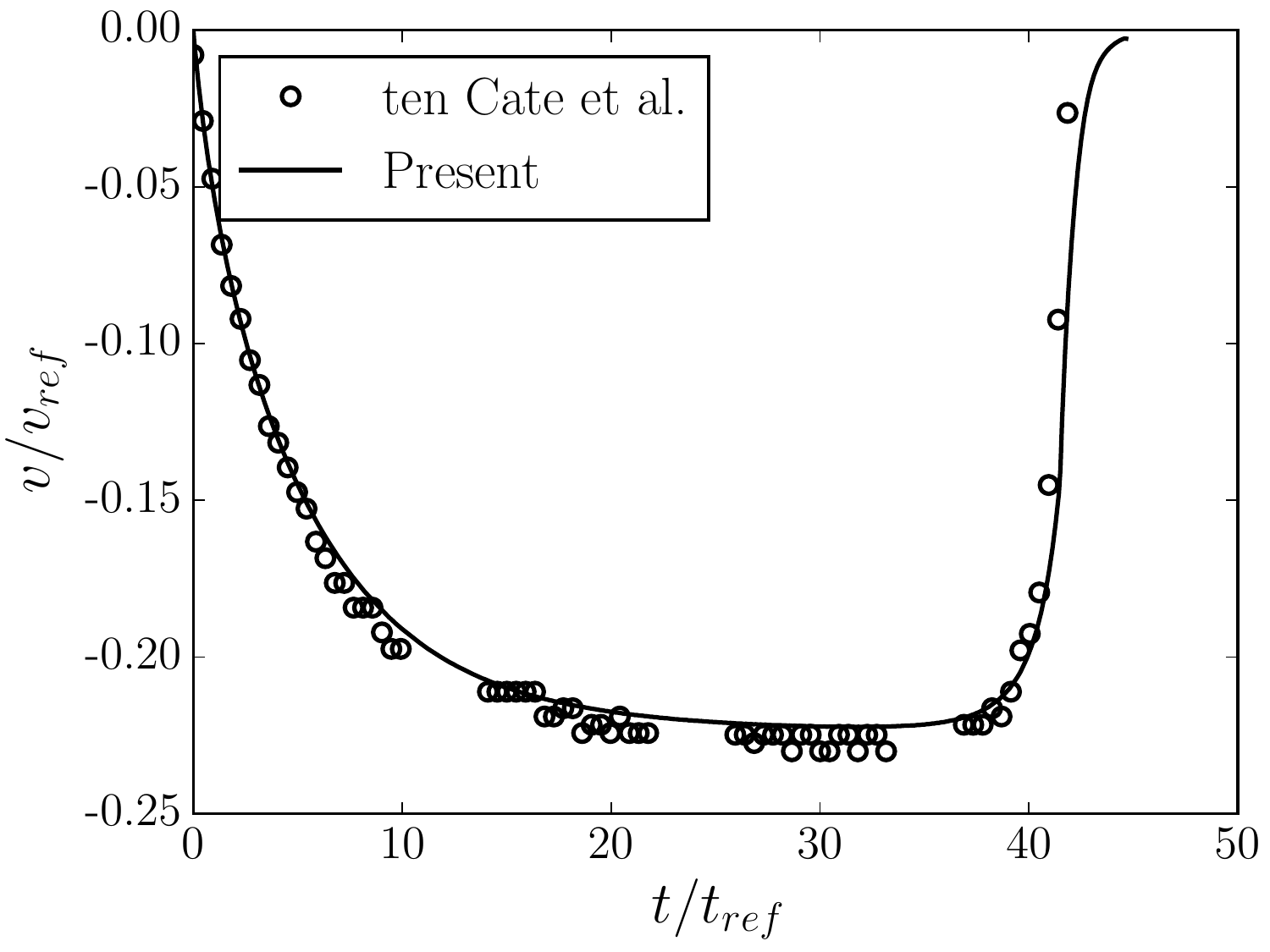}}
    \put( 0.0,  4.0)     {a) }
    \put( 6.5,  4.0)     {b) }
\end{picture}
        \caption{\small \textit{Comparison of the present numerical method against experimental data with $v_{ref}=\sqrt{gD_p}$ and $t_{ref}=\sqrt{D_p/g}$. a) Settling sphere in an infinite medium at $\mrm{Re}_p = 41$ \citep{mordant2000} and b) settling sphere in a wall-bounded medium at $\mrm{Re}_p = 12$ \citep{tencate2002}.}}
    \label{fig:settling_validation}
\end{figure}
%

\subsection{Structure of the collision model}\label{sec:coll_structure}
%
\setlength{\unitlength}{1cm}
\begin{figure}[t]
\begin{picture}(7,4.8)
  \put(0,  -0){\includegraphics[width=0.5\textwidth]{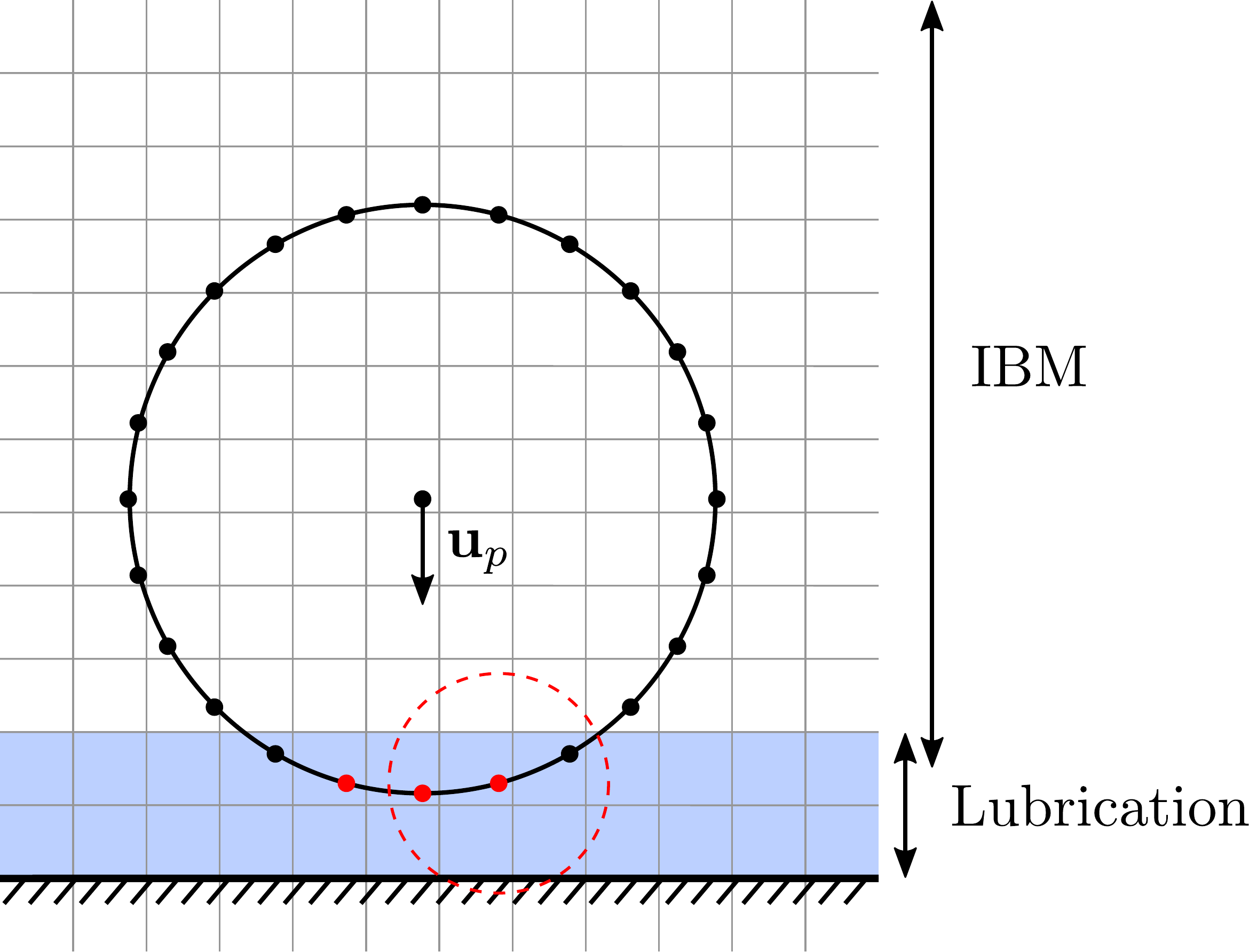}}
  \put(6.5,-0){\includegraphics[width=0.5\textwidth]{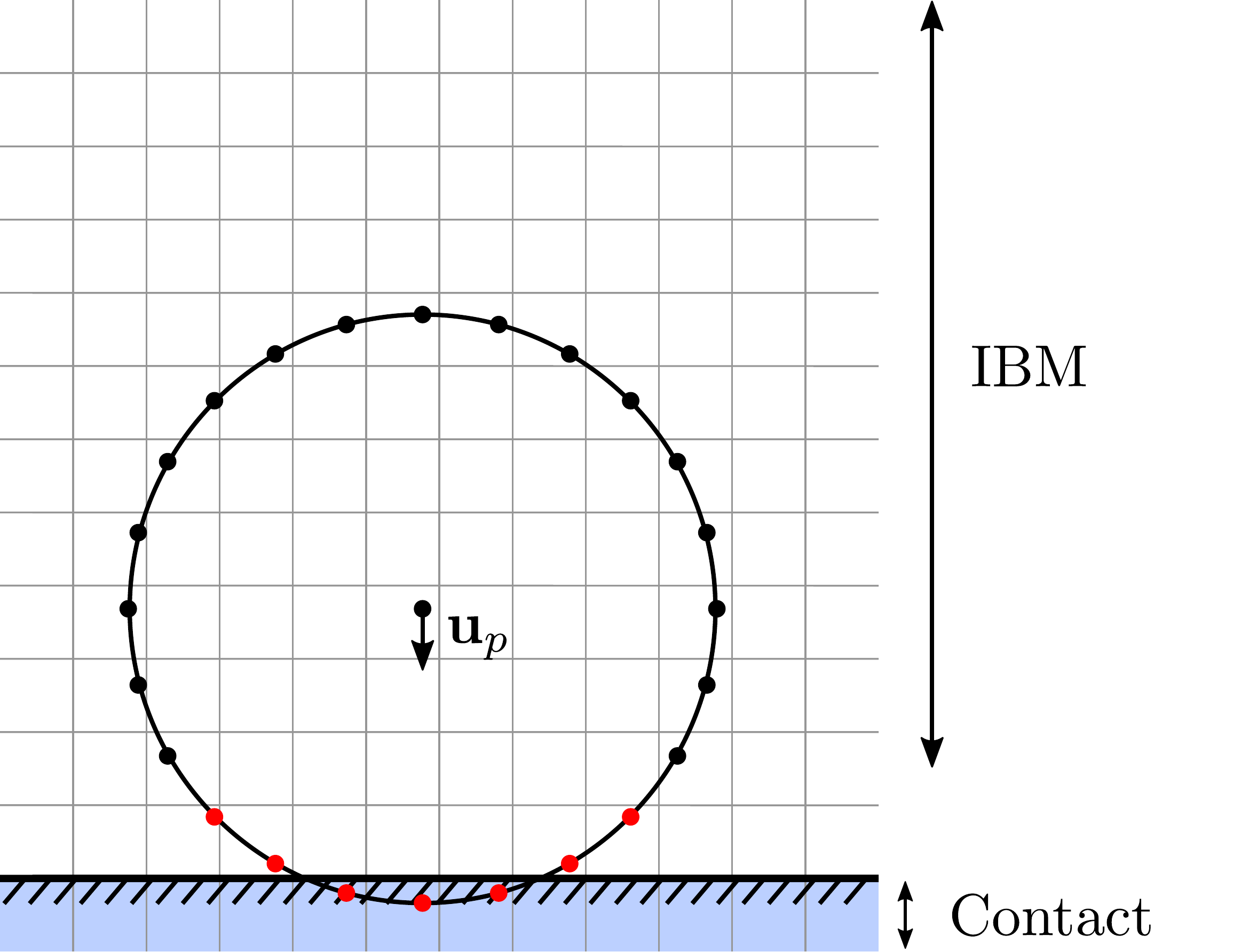}}
    \put( 0.0,  4.5)     {a) }
    \put( 6.5,  4.5)     {b) }
\end{picture}
        \caption{\small \textit{Regions where short-range interactions occur. Points on particle surface represent Lagrangian markers.  Red markers have been turned off. a) Lubrication region where red dashed circle illustrates support of Dirac delta function and b) Contact region.}}
    \label{fig:wall_collision}
\end{figure}
As mentioned in the introduction, one of the major advantages of the IBM is the direct computation of long-range interactions between the particles. Only short-range interactions and collisions need to be modeled.  For example, consider a particle approaching and colliding with a wall, as shown in Figure~\ref{fig:wall_collision}. As the particle comes close to the wall, two problems need to be dealt with: first, the smoothed Dirac delta functions used for the IBM overlap with the wall, and second, the discrete mesh can no longer resolve the fluid being squeezed out from between the two surfaces.

We can solve the first problem by disabling Lagrangian marker points whose supports overlap with the wall (red dashed circle in Figure~\ref{fig:wall_collision}a), as was done by \cite{kempe2012}.  This means that the forcing by these select markers on both the fluid and the particles is ignored, preventing the particle from using undefined information from outside of the domain and from competing with the wall for enforcing the no-slip condition.  Figure~\ref{fig:wall_collision} illustrates the red markers that have been disabled. Similarly, overlapping markers between two particles are disabled. We solve the second problem by adding a lubrication force, which models the subgrid forces on the particle due to the narrow gap and also accounts for some of the fluid forces from the disabled Lagrangian markers.  We apply this force when the particle-wall distance is less than two grid cells ($0 < \zeta_n \leq 2h$), illustrated by the blue region in Figure~\ref{fig:wall_collision}a. Once the particle comes into contact with the wall ($\zeta_n \leq 0$), we apply a contact force to prevent particles from overlapping too much and to account for proper momentum transfer and energy loss. This contact force involves components both normal and tangent to the two surfaces, representing material deformations and friction, respectively.

Hence, the following case distinctions can be made for the normal collision forces
\begin{linenomath*}
\begin{equation} \label{eq:coll_norm_conditions}
\mathbf{F}_n = \begin{cases}
	0 & \zeta_n > 2h \\
	\text{lubrication model \eqref{eq:lub_force}} & 0 < \zeta_n \leq 2h \\
	\text{normal contact model \eqref{eq:acm_force}} & \zeta_n \leq 0
	\end{cases}
\end{equation}
\end{linenomath*}
and the tangential collision forces
\begin{linenomath*}
\begin{equation} \label{eq:coll_tan_conditions}
\mathbf{F}_t = \begin{cases}
	0 & \zeta_n > 0 \\
	\text{tangential contact model \eqref{eq:lin_tan_coulomb}} & \zeta_n \leq 0 \qquad .
	\end{cases}
\end{equation}
\end{linenomath*}

\setlength{\unitlength}{1cm}
\begin{figure}[t]
\centering
\includegraphics[width=0.9\textwidth]{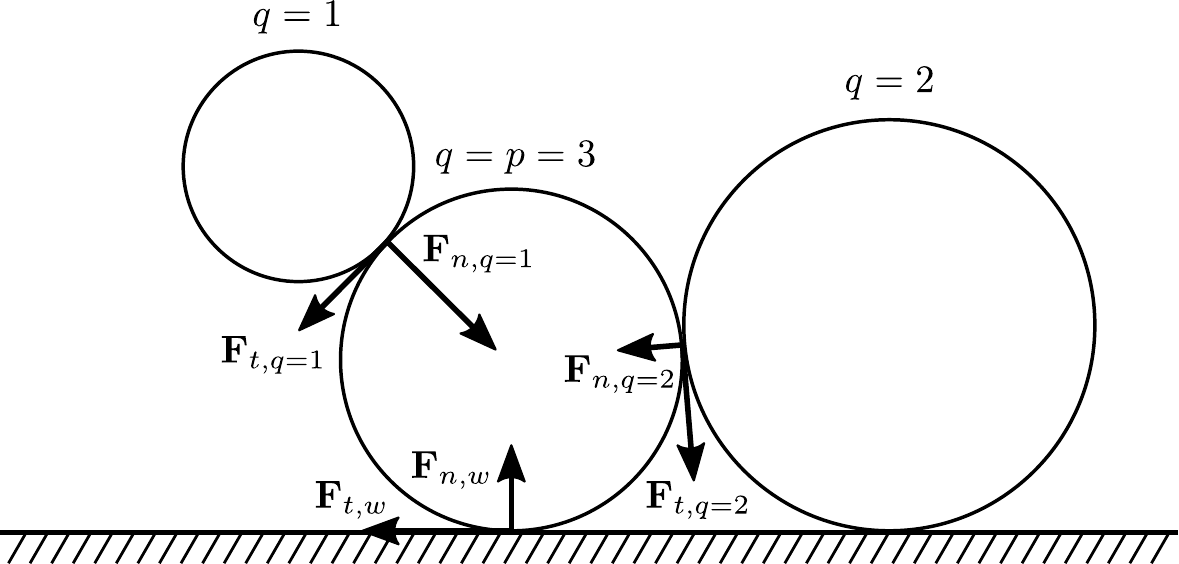}
\caption{\small \textit{Sketch of polydisperse particles in a mobile granular bed and the resulting interactions due to collisions.}}
\label{fig:granular_bed}
\end{figure}

The interactions of a single particle with its environment, however, are in general more diverse.
Let us consider a particle $p$ embedded in a mobile granular bed of polydisperse, spherical particles (Figure \ref{fig:granular_bed}). The dynamics of this particle are mainly determined by all the collision forces exerted upon it by particles $q, q \ne p$ as well as the wall. The total force $\textbf{F}_{c,p}$ acting on a particle $p$ during the collision process may be decomposed as
\begin{linenomath*}
\begin{equation}\label{eq:particle_forces}
	\textbf{F}_{c,p} = \sum_{q,\:q \neq p}^{N_p} \left( \textbf{F}_{n,q} + \textbf{F}_{t,q} \right) + \textbf{F}_{n,w} + \textbf{F}_{t,w}\hspace{0.5cm},
\end{equation}
\end{linenomath*}
where $N_p$ is the number of particles simulated, $\textbf{F}_{n,q}$ and $\textbf{F}_{n,w}$ are the normal collision forces described by \eqref{eq:coll_norm_conditions} with particle $q$ and the wall, respectively, and $\textbf{F}_{t,q}$ and $\textbf{F}_{t,w}$ are the tangential collision forces described by \eqref{eq:coll_tan_conditions} with particle $q$ and the wall, respectively. In what follows, the mathematical expressions are formulated for both particle-wall and particle-particle collisions, where the radii of the two colliding particles can be arbitrary. Whenever a distinction between particle-wall and particle-particle has to be made, a reference to \ref{app:definitions} is given for brevity, providing all definitions and nomenclature needed to distinguish between the two different cases.

The torque $\textbf{T}_{c,p}$ on a spherical particle $p$ generated by the tangential contact forces is
\begin{linenomath*}
\begin{equation}\label{eq:particle_moments}
	  \textbf{T}_{c,p} = \sum_{q,\:q \neq p}^{N_p} R_{p,cp} \:\textbf{n}_{\:p,q} \times  \textbf{F}_{t,q} + R_{p,cp} \:\textbf{n}_{\:p,w} \times  \textbf{F}_{t,w}
\end{equation}
\end{linenomath*}
where $\textbf{n}_{\:p,q}$ and $\textbf{n}_{\:p,w}$ are the unit vectors pointing to the collision partner $q$ or the wall, respectively,
and $R_{p,cp}$ is the particle radius at the contact point as defined per \ref{app:definitions} in \eqref{eq:R_cp}, which accounts for surface overlap.  In the next section, we will provide the mathematical description of the models used in the present study.

\section{Collision modeling}\label{sec:collision_modeling}

\subsection{Lubrication model}\label{sec:lubrication}
When the distance between the surfaces of two approaching particles becomes small, the fluid is squeezed out of the gap.
The fluid grid cannot resolve this process as soon as $\zeta_n < 2h$, where $h$ is the grid cell size.  Hence, we employ a lubrication model, which also acts on particles rebounding after the collision, when fluid is drawn into the gap.
The lubrication force is dissipative, since it is always directed opposite to the relative velocity.
The model is based on the analytical derivation of \cite{cox1967}, who solved for the force under Stokes flow conditions
\begin{linenomath*}
\begin{equation}\label{eq:lub_force}
\mbf{F}_n = - \frac{6 \pi \rho_f \nu_f R_\textit{eff}^2}{\max\left(\zeta_n, \zeta_{n,min} \right)} \mbf{g}_{n,cp} \qquad ,
\end{equation}
\end{linenomath*}
where $R_\textit{eff}$ is the effective radius accounting for polydisperse sediment and $\mbf{g}_{n,cp}$ the normal component of the relative particle velocity as defined per \ref{app:definitions} in \eqref{eq:R_eff} and \eqref{eq:g_ncp} respectively. The original model scales as $1/\zeta_n$ which introduces a singularity as $\zeta_n \to 0$. This has been addressed by \cite{simeonov2012} and \cite{kempe2012a}, who set $\mbf{F}_n = 0$ for $0<\zeta_n<\zeta_{n,min}$, and \cite{izard2014}, who have shifted the denominator of \eqref{eq:lub_force} to $\zeta_n + \zeta_{n,min}$. In the present approach, as in the approach of \cite{costa2015}, lubrication forces are held constant as soon as the gap size becomes smaller than the critical value $\zeta_{n,min}$, which provides a continuous forcing on the particles when they are close to sustained contact. This parameter can be interpreted as the micro-texture of the particle surface, which acts as a surface roughness, as will be discussed further in Section \ref{sec:surface_roughness} below.

\subsection{Normal contact model}\label{sec:coll_norm}
To account for normal contact forces, we implemented  the Adaptive Collision Time Model (ACTM) proposed by \cite{kempe2012a}. The main idea of the ACTM is to use an adaptive procedure to obtain the desired restitution coefficient $e_{dry}$ and to resolve the collision on the timescale of the fluid solver.  The ACTM is based on a nonlinear spring-dashpot system
\begin{linenomath*}
\begin{equation} \label{eq:acm_force}
 \mbf{F}_n = -k_n |\zeta_n|^{3/2} \mbf{n} - d_n \mbf{g}_{n,cp} \qquad ,
\end{equation}
\end{linenomath*}
which involves empirical parameters for the coefficients of stiffness $k_n$ and damping $d_n$. Here, $\mbf{n}$ is the normal vector pointing either towards the collision partner or towards the wall as defined per \ref{app:definitions} in \eqref{eq:n}. The nonlinear term $|\zeta_n|^{3/2}$ arises from Hertzian contact theory \citep{hertz1882}.

Since the timescale of a collision according to Hertzian contact theory is several orders of magnitude smaller than the typical temporal discretization of the fluid solver, the collision event has to be stretched in time to maintain the efficiency of the numerical procedure. This measure is also needed for the fluid to adapt to the sudden change in the particle trajectory \citep{kempe2012a, costa2015}.  However, the duration of contact $T_c$ is mainly determined by the stiffness parameter $k_n$. This becomes of particular importance for the complex situation of bed-load transport, where a broad range of impact velocities is encountered, ranging from high-impact collisions at the top of the bed to enduring contact within the bed.

The ACTM fixes this problem by adaptively calibrating the parameters $k_n$ and $d_n$ depending on the impact velocity $u_{in}$, the desired restitution coefficient $e_{dry}$, and the desired collision time $T_c$. The latter is a parameter of the model and should be minimized to avoid excessive particle overlaps and temporal stretching. \cite{kempe2012a} demonstrated that $T_c = 10 \Delta t$ is a suitable choice for the collision time given that all timescales related to fluid and particle motion are significantly larger than the timescale of particle contact.  For glass and hard metals, $e_{dry} = 0.97$ is a typical value \citep[e.g][]{foerster1994, joseph2001, gondret2002}. For immersed collisions, the restitution coefficient $e_{wet}$, measured some small distance away from the wall, becomes a function of the Stokes number \citep{joseph2001}.  The ACTM, however, uses the IBM and lubrication model to account for $e_{wet}$ through additional dissipative fluid effects.

In order to find values for $k_n$ and $d_n$, we first neglect all non-contact forces acting on the particle so that \eqref{eq:part_lin} and \eqref{eq:acm_force} together give the following nonlinear ordinary differential equation:
\begin{linenomath*}
\begin{equation}\label{eq:actm_ode}
 m_\textit{eff}\frac{\text{d}^2 \zeta_n}{\text{d}t^2}+d_n\frac{\text{d}\zeta_n}{\text{d}t}+ k_n \zeta_n^{3/2}=0 \qquad ,
\end{equation}
\end{linenomath*}
where $m_\textit{eff}$ is the effective mass  accounting for polydisperse sediment  as defined per \ref{app:definitions} in \eqref{eq:m_eff}.
Note that $\text{d}\zeta_n / \text{d}t = -\mbf{g}_{n,cp} \cdot \mbf{n}$. Together with \eqref{eq:actm_ode} and initial and final conditions, the constraints $e_{dry}$ and $T_c$ allow for determination of $k_n$ and $d_n$ using either an iterative procedure, as was done by \cite{kempe2012a}, or an explicit formulation, as was proposed by \cite{ray2015}.
In the present study, we implemented the explicit formulation, which is provided in \ref{app:ACTM_coeff}.
According to \cite{ray2015}, the error in $u_{out}$ increases with decreasing $e_{dry}$, but does not exceed 1.3\% for $e_{dry} > 0.7$ or 3\% for $e_{dry} > 0.4$, making this method useful for most sediment materials such as silicate, glass, or even metal.

\subsection{Tangential collision model}\label{sec:tangential_force}
%
To account for frictional contact between the particles, we implemented a tangential contact model based on the linear spring-dashpot model described in the review paper of \cite{thornton2013}:
\begin{linenomath*}
\begin{equation} \label{eq:lin_tan}
\mbf{F}_{t,LS} = -k_t \mbs{\zeta}_t - d_t \mbf{g}_{t,cp}  \qquad ,
\end{equation}
\end{linenomath*}
which has stiffness and damping coefficients $k_t$ and $d_t$.  This model uses $\mbf{g}_{t,cp}$, the tangential component of the relative surface velocities as described in \eqref{eq:g_tcp} of \ref{app:definitions}, as well as $\mbs{\zeta}_t$, the tangential spring displacement, which represents the accumulated relative tangential motion between the two surfaces:
\begin{linenomath*}
\begin{equation} \label{eq:zeta_t_int}
\mbs{\zeta}_t = \int_{t_i}^t \mbf{g}_{t,cp}(t') dt'  \qquad ,
\end{equation}
\end{linenomath*}
where $t_i$ is the time of impact.  The discretized form of \eqref{eq:zeta_t_int} is described in \ref{app:zeta_t}.

This model limits the maximum force based on Coulomb's friction criterion:
\begin{linenomath*}
\begin{equation} \label{eq:lin_tan_coulomb}
\mbf{F}_t = \min \left(||\mbf{F}_{t,LS}||, ||\mu \mbf{F}_n||\right) \mbf{t} \qquad ,
\end{equation}
\end{linenomath*}
where $\mu$ represents the coefficient of friction between the two surfaces (described further in Section~\ref{sec:rolling_sliding}) and $\mbf{t} = \mbf{F}_{t,LS} / ||\mbf{F}_{t,LS}||$ points in the direction of the tangential force.

This model has two important features for simulating densely-packed beds.  First, the spring allows many particles to interact in a smooth, stable manner, provided the stiffness is chosen properly. Second, the model has a memory of the friction force via the tangential displacement $\mbs{\zeta}_t$, which permits a steady-state frictional bed configuration.  In contrast, a model that only uses $\mbf{g}_{t,cp}$, such as the one proposed by \cite{kempe2012a}, can only react to slip, not predict it.

Similarly to the ACTM, we can adaptively compute $k_t$ and $d_t$ for each collision.  According to \cite{thornton2011}, the stiffness can be set to
\begin{linenomath*}
\begin{equation} \label{eq:k_t}
k_t = \frac{\kappa \, m_\textit{eff} \, \pi^2}{T_c^2} \qquad .
\end{equation}
\end{linenomath*}
Here, $\kappa$ is based on Poisson's ratio $\nu$:
\begin{linenomath*}
\begin{equation} \label{eq:kappa}
\kappa = \frac{2(1-\nu)}{2-\nu} \qquad ,
\end{equation}
\end{linenomath*}
which is a well-studied material property typically ranging between $0.22<\nu<0.30$ \citep[e.g][]{foerster1994,gondret2002,joseph2004}. Hence, a value of $\nu = 0.22$ was used in the present study.

In addition, the damping is computed according to \cite{thornton2013} to account for the inelasticity of the collisions
\begin{linenomath*}
\begin{equation} \label{eq:d_t}
d_t = 2\sqrt{ m_\textit{eff}\, k_t }\frac{-\mrm{ln}\,e_{dry}}{\sqrt{\pi^2 + \mrm{ln}^2e_{dry}}} \qquad .
\end{equation}
\end{linenomath*}
Having created a uniform collision time $T_c$ with the normal contact model, we obtain the correct rebound characteristics for oblique impacts using these values for $k_t$ and $d_t$, as shown in Section~\ref{sec:rolling_sliding}. Consistently, the model does not require any calibration but instead can be parameterized using material properties obtained from experiments.
%
\section{Enhancements to the normal contact model}\label{sec:enhancements}
%
\subsection{Motivation}
%
In order to obtain a good agreement with immersed collision experiments, we had to implement a few enhancements to the normal contact model described in Section~\ref{sec:coll_norm}.  Both changing the time integration to a scheme of higher accuracy and adding more timesteps to the integration of particle motion without changing the fluid timestep allowed us to reproduce the collision trajectories of \cite{gondret2002} in a robust manner.

\subsection{Improved time integration}\label{sec:time_integration}

The ACTM normal contact force $\mbf{F}_{c,p}$ is a function of the surface distance $\zeta_n$ and the relative velocity $\mbf{g}_{n,cp}$, which in turn depend on the particle position $\mbs{x}_p^{k-1}$ and velocity $\mbs{u}_p^{k-1}$ at the previous substep $k-1$.  We can write this functional dependence as $\mbf{F}_{c,p}(\mbf{x}_p^{k-1}, \mbf{u}_p^{k-1})$.  Integrating the particle equation of motion with a Forward Euler/Crank Nicholson scheme for the particle's velocity/position, we obtain:
\begin{IEEEeqnarray}{rCl}\label{eq:forward_euler}\IEEEyesnumber\IEEEyessubnumber*
\mbf{u}_p^k &=& \mbf{u}_p^{k-1} + \frac{2 \Delta t \alpha_k}{m_p}
	\mbf{F}_{c,p} \left( \mbf{x}_p^{k-1}, \, \mbf{u}_p^{k-1} \right) \\
\mbf{x}_p^k &=& \mbf{x}_p^{k-1} + \Delta t \alpha_k
	\left( \mbf{u}_p^k + \mbf{u}_p^{k-1} \right) \qquad ,
\end{IEEEeqnarray}
where $k$ is the number of the RK-substep and $\alpha_k$ is the RK-coefficient \citep{rai1991}. For now, we ignore the hydrodynamic, gravitational, and lubrication forces in order to focus on the contact forces alone.  We conducted a simple test to analyze the accuracy of this scheme.  A particle of density $\rho_p / \rho_f = 7.8 $ and with radius $R_p=10 u_\infty \Delta t$ was initialized with a velocity of $\textbf{u}_p = \left(0, u_\infty , 0\right)^T$ at a position $y_p > R_p$ above a wall at $y=0$. Subsequently, the particle was released and eventually collided with the wall. Neglecting hydrodynamic effects as well as gravity yields an impact velocity of $u_{in} / u_\infty=1 $. Choosing the collision time to be $T_c = 10 \Delta t$ as suggested by \cite{kempe2012a}, gave good results for the  duration of the desired contact phase $T_c$, but rather large errors of the rebound velocity were observed compared to the prescribed $e_{dry} = 1$. The value of $u_{out} = -e_{dry} \, u_{in}$ was overestimated by more than $12 \%$.

\begin{figure}[t]
\centering
\includegraphics[width=0.65\textwidth]{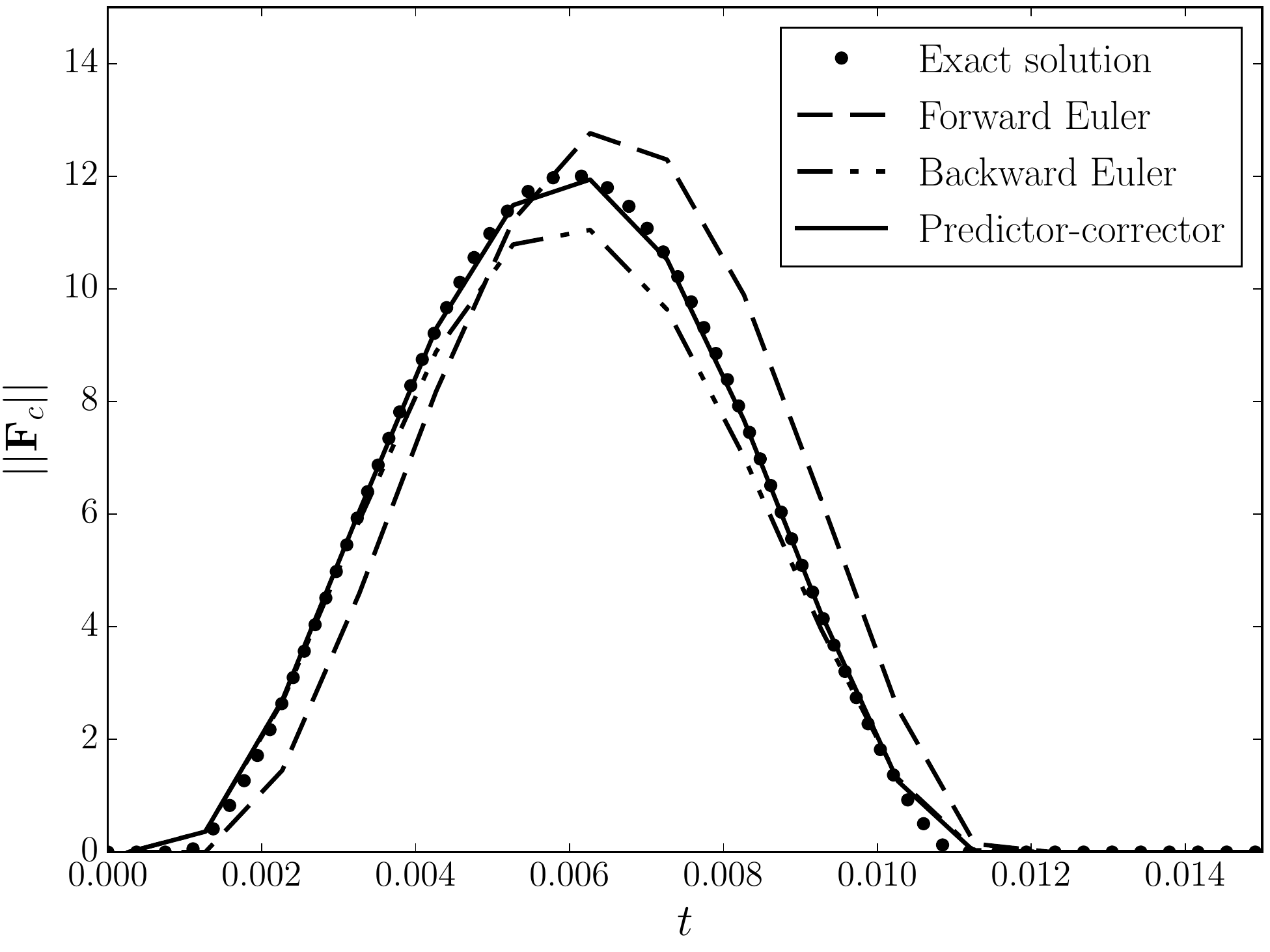}
\caption{\small \textit{Collision forces vs. time for Forward Euler, Backward Euler, and predictor-corrector schemes.}}
\label{fig:Fc_vs_t_mod}
\end{figure}

Turning our attention to Figure~\ref{fig:Fc_vs_t_mod}, we can see that the discretization of \eqref{eq:forward_euler} leads to a poor estimation of the collision force $||\mbf{F}_{c,p}||$ when compared to the simulation in which $10^4$ timesteps were used to resolve the collision, which can be taken as the exact solution.  This inaccuracy in the collision force was observed for a variety of simulations using different $R_p$, $\rho_p$, $u_{in}$, and $e_{dry}$.  In order to reduce the error to $0.1$\%, a temporal discretization of $T_c = 1000 \Delta t$ would be required, which is not feasible for simulations of sediment transport.  Hence, we implemented a temporal discretization scheme with a higher order of accuracy. Utilizing the same three-step RK scheme that integrates the Navier-Stokes equations, we reformulated the collision integration with a predictor-corrector scheme:
\begin{IEEEeqnarray}{rCl}\label{eq:predictor_corrector} \IEEEyesnumber\IEEEyessubnumber*
\widetilde{\mbf{u}}_p &=& \mbf{u}_p^{k-1} + \frac{\Delta t}{m_p}
	\left[ \gamma_k \mbf{F}_{c,p} \left( \mbf{x}_p^{k-1}, \, \mbf{u}_p^{k-1} \right)
	+ \zeta_k \mbf{F}_{c,p} \left( \mbf{x}_p^{k-2}, \, \mbf{u}_p^{k-2} \right) \right]
	\label{eq_ucoll_predict} \\
\widetilde{\mbf{x}}_p &=& \mbf{x}_p^{k-1} + \Delta t \, \alpha_k
	\left( \widetilde{\mbf{u}}_p + \mbf{u}_p^{k-1} \right) \\
\mbf{u}_p^k &=& \mbf{u}_p^{k-1} + \frac{\Delta t \, \alpha_k}{m_p}
	\left[ \mbf{F}_{c,p} \left( \widetilde{\mbf{x}}_p, \, \widetilde{\mbf{u}}_p \right)
	+  \mbf{F}_{c,p} \left( \mbf{x}_p^{k-1}, \, \mbf{u}_p^{k-1} \right) \right] \\
\mbf{x}_p^k &=& \mbf{x}_p^{k-1} + \Delta t \, \alpha_k
	\left( \mbf{u}_p^k + \mbf{u}_p^{k-1} \right) \qquad .
\end{IEEEeqnarray}
Here, tilde indicates predicted values, and $\gamma_k$ and $\zeta_k$ are the RK coefficients for the explicit third-order scheme according to \cite{rai1991}. Hence, the velocity predictor step \eqref{eq_ucoll_predict} is third-order accurate while the other steps use second-order Crank-Nicholson schemes. A similar approach was taken by \cite{costa2015}, but in this reference, the predicted value is determined by an iterative scheme, which is computationally more costly than the present scheme. In Figure~\ref{fig:Fc_vs_t_mod} we can see that this predictor-corrector scheme yields a much better approximation of $||\mbf{F}_{c,p}||$ compared to the Forward Euler method, reducing the error of $u_{out}$ by almost two orders of magnitude to $0.17 \%$ for $T_c/\Delta t = 10$.  For completeness, we have also included a solution that uses the Backward Euler method, which underestimates the rebound velocity by 11\%. This improvement has been achieved by a minimal increase of the computational costs, as the most expensive part of \eqref{eq:part_lin} is the computation of the hydrodynamic forces $\textbf{F}_{h,p}$.
\setlength{\unitlength}{1cm}

\subsection{Temporal substepping} \label{sec_substep}

\begin{table}[t]
\centering
\begin{tabular}{lcc}
\hline
\hline
$\mrm{St}$			& 27					& 152		\\
$\mrm{Re}_p$		& 30					& 164		\\
\hline
$R_p$ (m)			& 0.003					& 0.0015	\\
$u_{in}$ (m/s)		& 0.518					& 0.585		\\
$\rho_p / \rho_f$	& 8.083					& 8.342		\\
$\nu_f$ (m$^2$/s)	& $1.036\times10^{-4}$	& $1.070\times10^{-5}$	\\
$e_{dry}$			& 0.97					& 0.97		\\
$g$ (m/s$^2$)		& 9.81					& 9.81		\\
\hline
Domain size (m) ($L_x \times L_y \times L_z$)
	& $0.08 \times 0.16 \times 0.08$	& $0.02 \times 0.2 \times 0.02$	\\
Domain boundary conditions
	& p $\times$ ns $\times$ p			& p $\times$ ns $\times$ p		\\
Initial position of sphere center (m)	& 0.075		& 0.197				\\
Grid cells in $x$-direction			& 256			& 128				\\
Grid cells per diameter				& 19			& 19				\\
Timestep			& $\Delta t = 2.5\text{e-4}$	& $\Delta t = 8.9\text{e-5}$ \\
\hline
\hline
\end{tabular}
\caption{\small \textit{Simulation parameters to match the experiments of \cite{gondret2002}.  Boundary conditions can be periodic (p), slip (s), or no-slip (ns).}}
\label{tab:gondret}
\end{table}
Having improved the accuracy of the contact model, we carried out simulations of particle-wall collisions in a fluid to compare to the experiments of \cite{gondret2002}. The details of the simulations, including the material properties as well as the physical and numerical parameters, are summarized in Table~\ref{tab:gondret}. \cite{gondret2002} released particles from heights large enough to accelerate to their terminal velocities before colliding with the wall.
For these simulations, the horizontal wall and vertical particle trajectories allow us to only consider normal collision forces.
To control the impact velocity $u_{in}$, we accelerated the particle in the numerical simulations according to the relation
\begin{equation}
u(t) = u_{in} \left( e^{-40t} -1 \right), \quad \zeta_n > R_p \qquad .
\end{equation}
In other words, we prescribed the falling velocity of the particle so that it accelerated in a smooth manner so that $u_{in}$ matched the Stokes number reported in \cite{gondret2002} as shown in Table~\ref{tab:gondret}.  Two scenarios were considered: one with a rather high Stokes number $\mrm{St}= 152$ and one with a lower Stokes number of $\mrm{St} = 27$, the latter of which is within the range of Stokes numbers that have been reported for the numerical simulations of \cite{kempe2014}. Once the particle reached a distance of $\zeta_n = R_p$, we turned off the prescribed velocity, allowing the particle to move on its own volition according to the hydrodynamic, buoyant, and collision forces acting on it.

While attempting to reproduce the experimental trajectories, the simulations produced large variations in the results from small changes to the initial conditions.
To show this, we executed five simulations for $\mrm{St}=27$, varying only the initial position of the particle $y_0$ from the value recorded in Table~\ref{tab:gondret} within the interval of one grid cell $h$.  Figure~\ref{fig:sensitivity}a shows the range of trajectories encountered. For $\mrm{St} = 27$, a substantial variation in the rebound height of up to 83\% can be observed.  We would thus expect the collision model to produce a variety of incorrect trajectories for the simulation of sediment transport in a horizontal channel flow. Even the mean of the variety of trajectories is not able to fully reproduce the experimental trajectory.

\setlength{\unitlength}{1cm}
\begin{figure}[t]
\begin{picture}(7,4.5)
  \put(0,  -0){\includegraphics[width=0.5\textwidth]{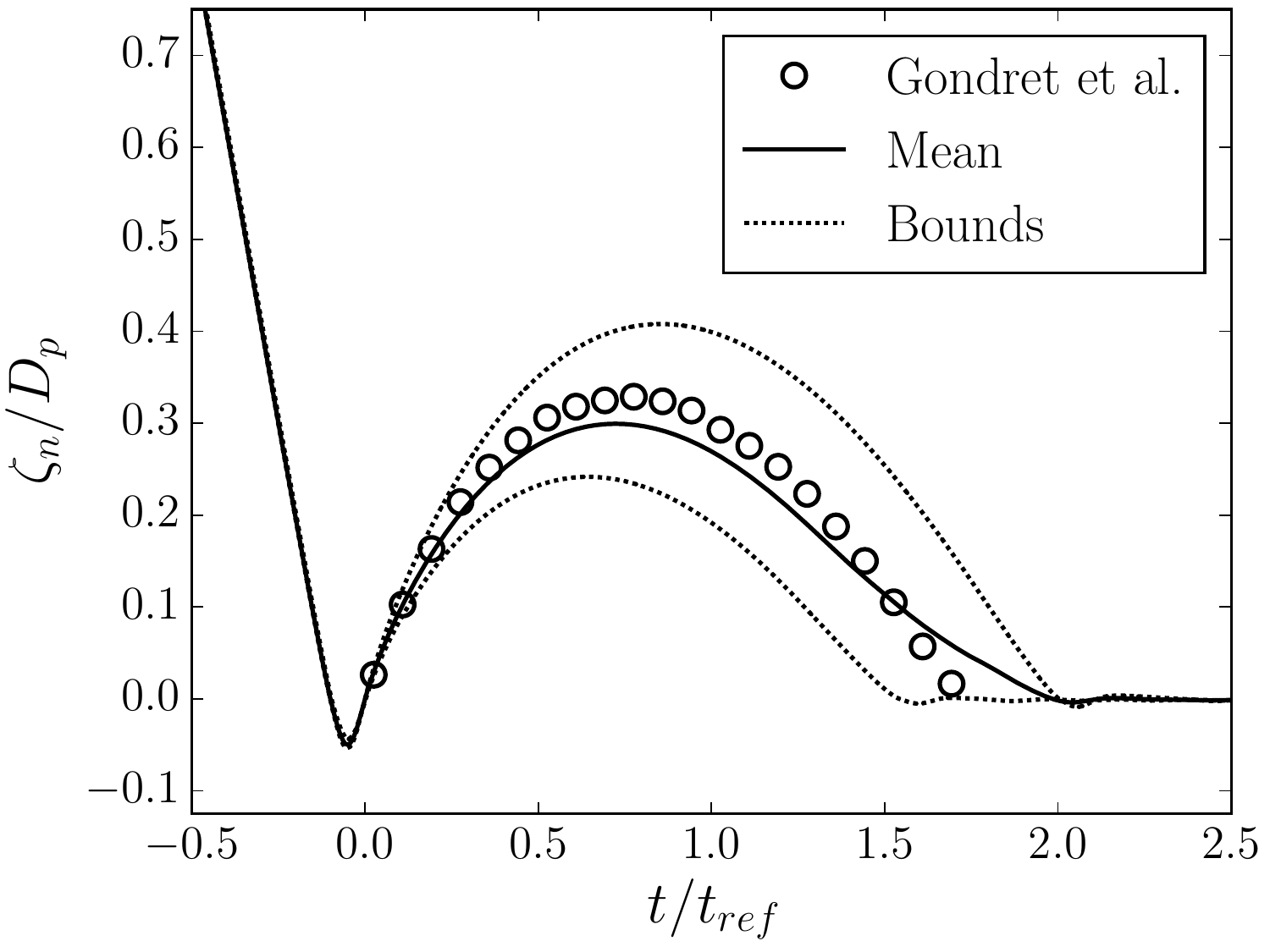}}
  \put(6.5,-0){\includegraphics[width=0.5\textwidth]{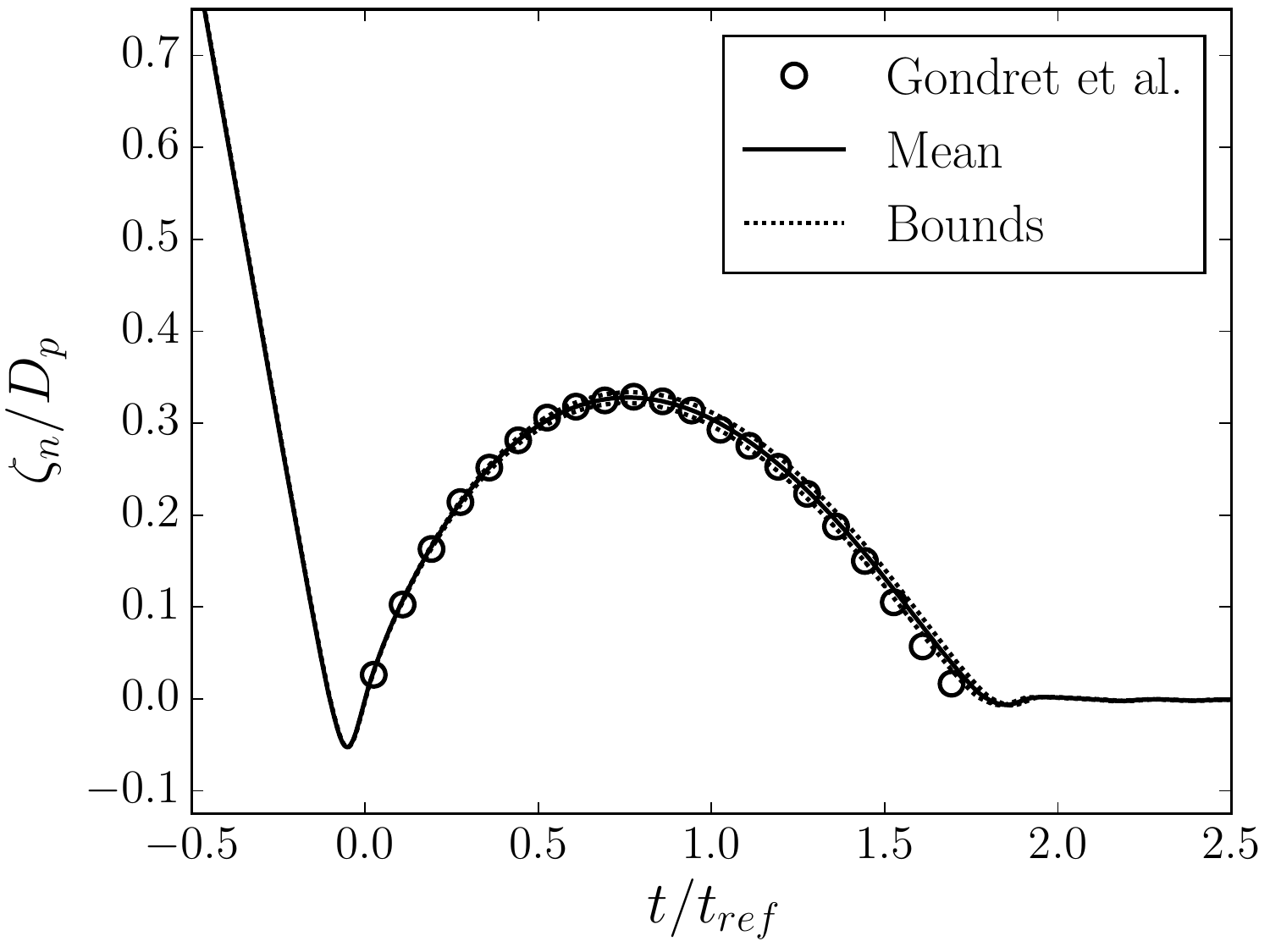}}
    \put( 0.0,  4.2)     {a) }
    \put( 6.5,  4.2)     {b) }
\end{picture}
        \caption{\small \textit{Sensitivity of rebound trajectories to initial position $y_0$ for $\mrm{St}=27$. a) Trajectories computed without particle substeps and b) trajectories computed with particle substeps.}}
    \label{fig:sensitivity}
\end{figure}
%

\begin{figure}[t]
\centering
\includegraphics[width=0.65\textwidth]{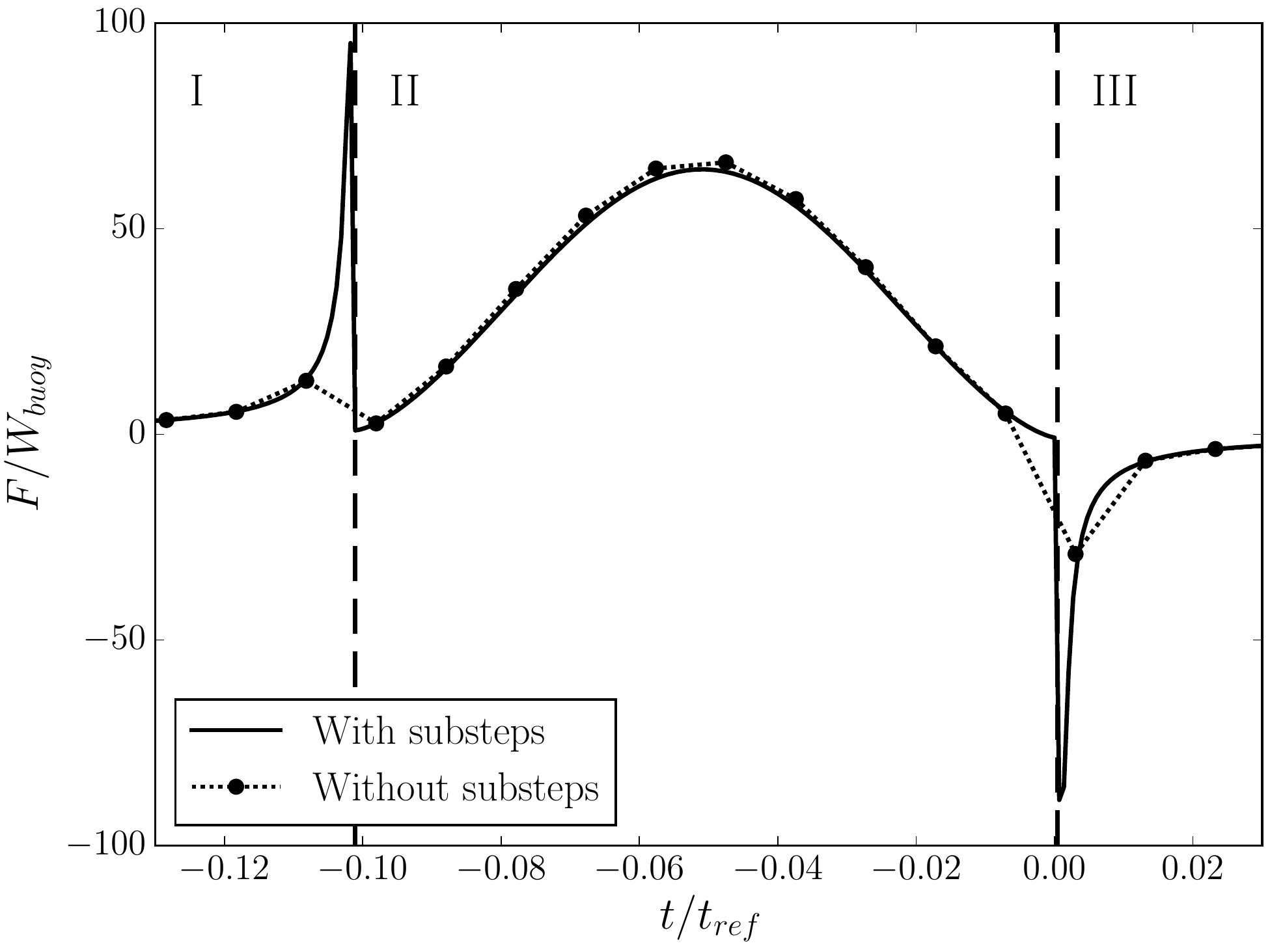}
\caption{\small \textit{Collision forces acting on the particle at $\mrm{St} = 27$ including: the lubrication force during approach (phase I), the normal force during contact (phase II), and the lubrication force during rebound (phase III).  Vertical dashed lines indicate a change in phase.  $W_{buoy}=(1-\rho_f/\rho_p)m_p \,g$ is the buoyant weight of the particle.}}
\label{fig:sensitivity_forces}
\end{figure}

To better understand the observed variability, we plot the time evolution of the collision forces, i.e. lubrication and contact forces, for the low Stokes number case $\mrm{St} = 27$ in Figure~\ref{fig:sensitivity_forces}. In this plot, we can see the particle approaching the wall with the lubrication force growing as $1/\zeta_n$ (phase I). Subsequently, the lubrication forces become zero during the contact phase starting at $t/t_{ref} = -0.1$. During this phase (phase II), the contact force grows and then decays with the particle-wall overlap as the particle changes direction to rebound. Finally, the particle experiences the lubrication force again during the rebound phase starting at $t/t_{ref} = 0$ (phase III). At this time, lubrication is acting in the opposite direction because lubrication is dissipative. The dotted line in Figure~\ref{fig:sensitivity_forces} shows the forces acting on the particle for a time discretization based on $\mrm{CFL}=0.5$ for the settling velocity. As expected, the normal contact model with the modifications described in Section \ref{sec:time_integration} above is able to give a smooth evolution of contact forces with the time step size of the fluid solver.  However, it turns out that the lubrication forces remain under-resolved during approach and rebound, especially as $\zeta_n$ approaches zero directly before and after the contact phase. This leads to either more or less total impulse acting on the particle, depending on where the timestep happens to land, which in turn results in variability between simulations. This effect strongly depends on the Stokes number, since the lubrication force decreases with increasing $\mrm{St}$. Hence, the ratio of the normal contact force to the lubrication force increases when approaching dry contact conditions.

Since the lubrication model used is an algebraic relation that does not depend on the surrounding hydrodynamics, we have implemented a substepping method that integrates the particle motion with smaller timesteps than the fluid motion. This method works as follows:
\begin{enumerate}
\item We solve the fluid equations of motion, IBM, and hydrodynamic forces acting on the particle as normal.

\item We divide the fluid RK substep $k$ into a number of substeps $N_{sub,k} = \{8, 2, 5\}$.  This choice results in a total of 15 substeps of constant size per fluid timestep ($\Delta t_{sub} = \Delta t / 15$), which is most efficient since $2 \alpha_k = \{8/15, 2/15, 5/15\}$ as used in \eqref{eq:predictor_corrector}.

\item For each of the substeps, we solve the particle equations of motion with the three-step RK method.  As we update the particle velocities and positions, we re-evaluate the collision (lubrication and contact) forces, but the hydrodynamic forces remain constant. This compromise makes the present approach very efficient.

\item At the end of the 8, 2, or 5 substeps, we use the final particle position and velocity for the next fluid RK substep.

\end{enumerate}
This measure effectively increases the resolution of a collision to a timestep 15 times smaller than the fluid timestep to integrate particle motion, allowing us to compute the lubrication forces with higher accuracy. Since the contact duration of $T_c = 10 \Delta t$ is maintained, the contact phase is now resolved with a total of 150 timesteps with only a marginal increase to the computational cost. Substepping has also been used by \cite{kidanemariam2014} and \cite{costa2015} but the authors did not illustrate the variability we have observed for the trajectories of particle-wall collisions. Meanwhile, \cite{kidanemariam2014} do not provide a comparison with the data of \cite{gondret2002} at all. The scheme presented by \cite{costa2015} still relies on an iterative procedure subdividing every fluid timestep into 50 substeps, which is less efficient than the scheme presented here. The results of our approach can be appreciated in Figure~\ref{fig:sensitivity_forces}. The solid line, which was resolved with fifteen times more timesteps, can be viewed as a better approximation of the exact solution to the model we have implemented. Figure~\ref{fig:sensitivity}b shows how this method almost eliminates the variability in the rebound trajectories of the particle-wall collisions discussed above.

\subsection{Choice of particle surface roughness}\label{sec:surface_roughness}

As shown in Figure~\ref{fig:sensitivity}, the improved integration scheme described in Sections \ref{sec:time_integration} and \ref{sec_substep} yields excellent results in reproducing the rebound trajectory of the $\mrm{St}=27$ experiment of \cite{gondret2002}.  Having obtained consistent results that are insensitive to the initial condition, we can use the same setup of particle-wall collisions to select the most suitable surface roughness $\zeta_{n,min}$ for the lubrication model \eqref{eq:lub_force}. This is the only parameter involved that requires calibration as an inverse problem. However, the range of values that can be assigned to $\zeta_{n,min}$ should neither fall below the surface roughness of the actual simulated particle nor exceed the length of a grid cell in order for the lubrication model to make physical sense.

\begin{figure}[t]
\centering
\includegraphics[width=0.65\textwidth]{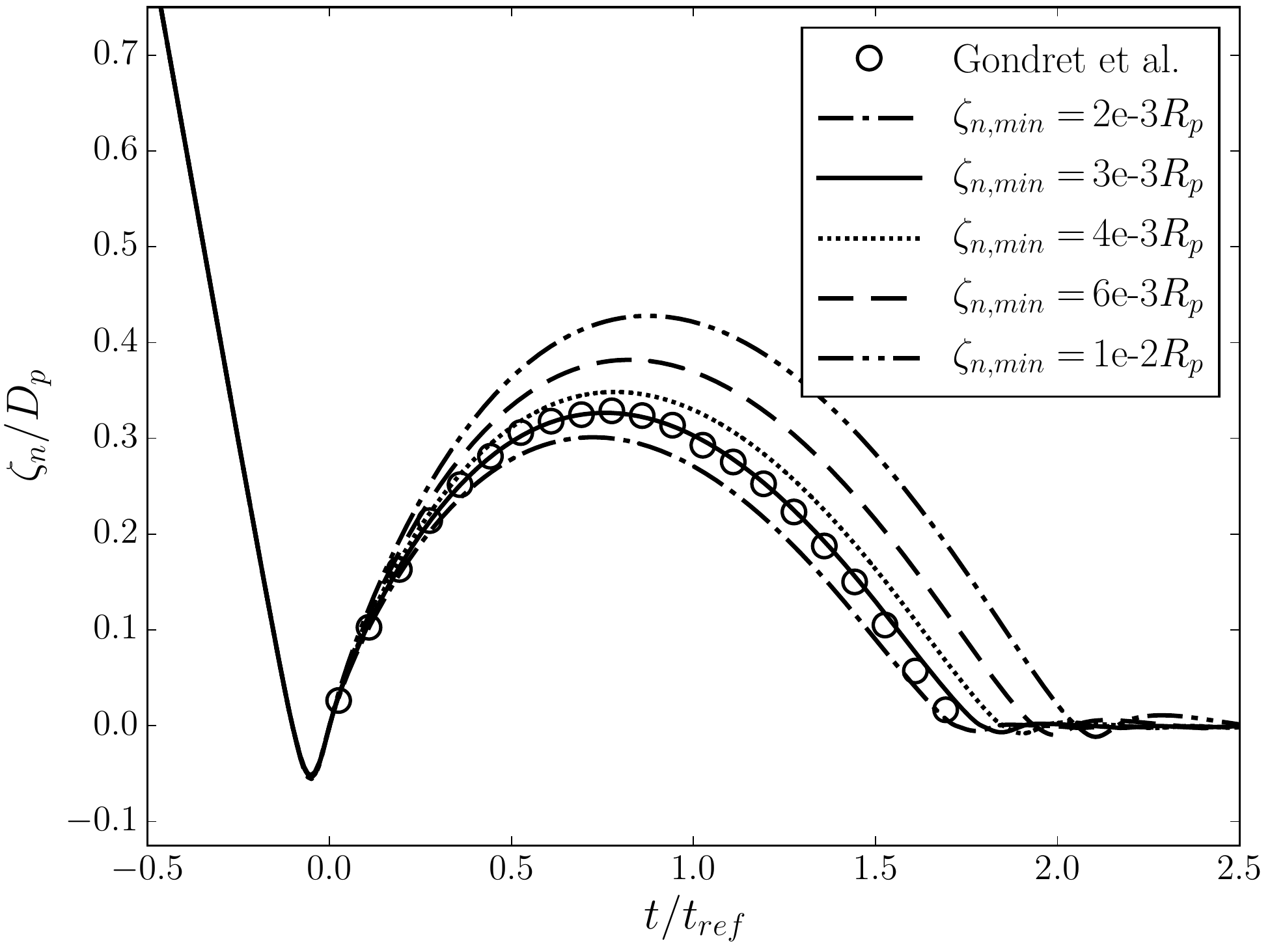}
\caption{\small \textit{Effect of changing $\zeta_{n,min}$ on rebound trajectories for $\mrm{St}=27$.}}
\label{fig:lub_cut}
\end{figure}

The impact of $\zeta_{n,min}$ on particle rebound trajectories for $\mrm{St}=27$ is illustrated in Figure~\ref{fig:lub_cut}. A clear trend can be identified: decreasing the value of $\zeta_{n,min}$ also decreases the rebound height due to more damping within the lubrication layer.  However, the results are moderately sensitive to the roughness value.  For instance, note that changing the roughness by an order of magnitude has a similar effect to excluding substeps (as shown in Figure~\ref{fig:sensitivity}).  Based on the present results, we selected $\zeta_{n,min} = 3\text{e-3}R_p$ to optimize agreement with the experimental data.  We have used this value for all simulations in the present work.  Note that a surface roughness of $1\text{e-4}R_p$ has been reported by \cite{gondret2002}, and other authors have used the physical particle roughness length for this parameter \citep{kempe2012a, costa2015} to avoid the singularity in the lubrication force.
Thus, we do not consider this parameter to be an exact physical representation of the actual surface roughness, but rather as a parameter to be calibrated within a reasonable range (small enough to be meaningful relative to the particle size and large enough to be resolved by the substeps).

\subsection{Particle momentum balance for high Stokes number collisions}\label{sec:neglecting_fluid_force}

Finally, we present a clarification to the ACTM as written by \cite{kempe2012a}. As already mentioned in Section \ref{sec:coll_norm}, the ACTM assumes that \eqref{eq:actm_ode} represents the equation of motion for the particle in determining the coefficients $k_n$ and $d_n$. In other words, no fluid or gravitational forces act on the particle during the contact phase.  Though not stated in their paper, \cite{kempe2012a} excluded hydrodynamic and buoyant weight forces in order to reproduce the trajectories of \cite{gondret2002} (Kempe \& Fr\"ohlich, 2016, private communication).  Thus, during contact the non-disabled Lagrangian markers still affect the fluid, but not the particle momentum.  This procedure is somewhat delicate for the situation of sediment transport in a horizontal channel considering the fact that the governing nondimensional number is the ratio of the hydrodynamic stress to the buoyant weight of the particle. This characteristic number is classically known as the Shields parameter $\mrm{Sh} = \tau_w / ((\rho_p-\rho_f) g D_p)$, where $\tau_w$ is the wall shear stress \citep{shields1936}. It is, therefore, very desirable to include gravitational and buoyant forces in \eqref{eq:part_lin} during the contact phase. In our experience, including the gravitational force during contact has a negligible effect in changing the desired $T_c$ and $e_{dry}$. In fact, the results presented so far have all been generated by including buoyant weight during contact.

%
%
\setlength{\unitlength}{1cm}
\begin{figure}[t]
\begin{picture}(7,4.5)
  \put(0,  -0){\includegraphics[width=0.5\textwidth]{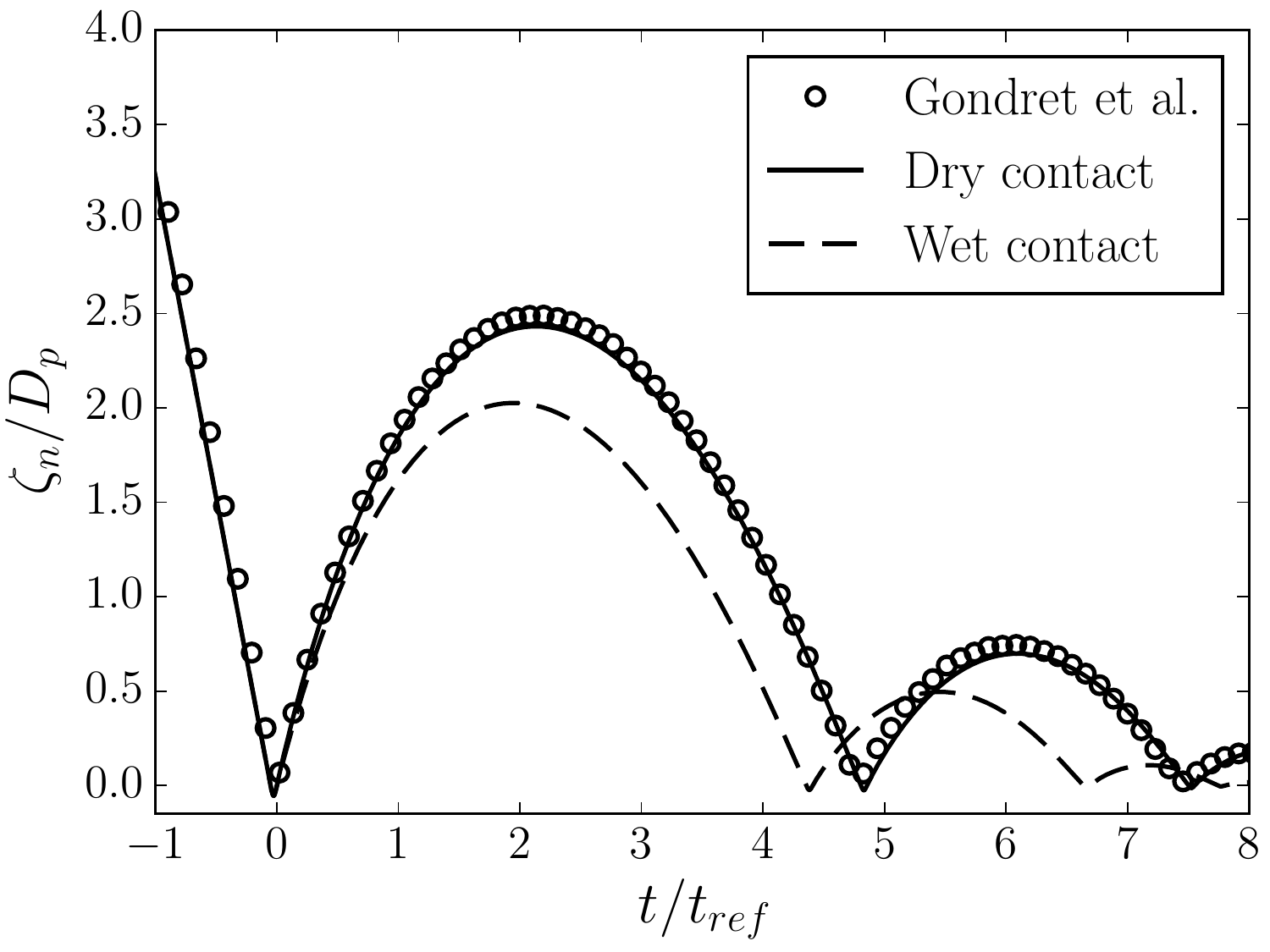}}
  \put(6.5,-0){\includegraphics[width=0.5\textwidth]{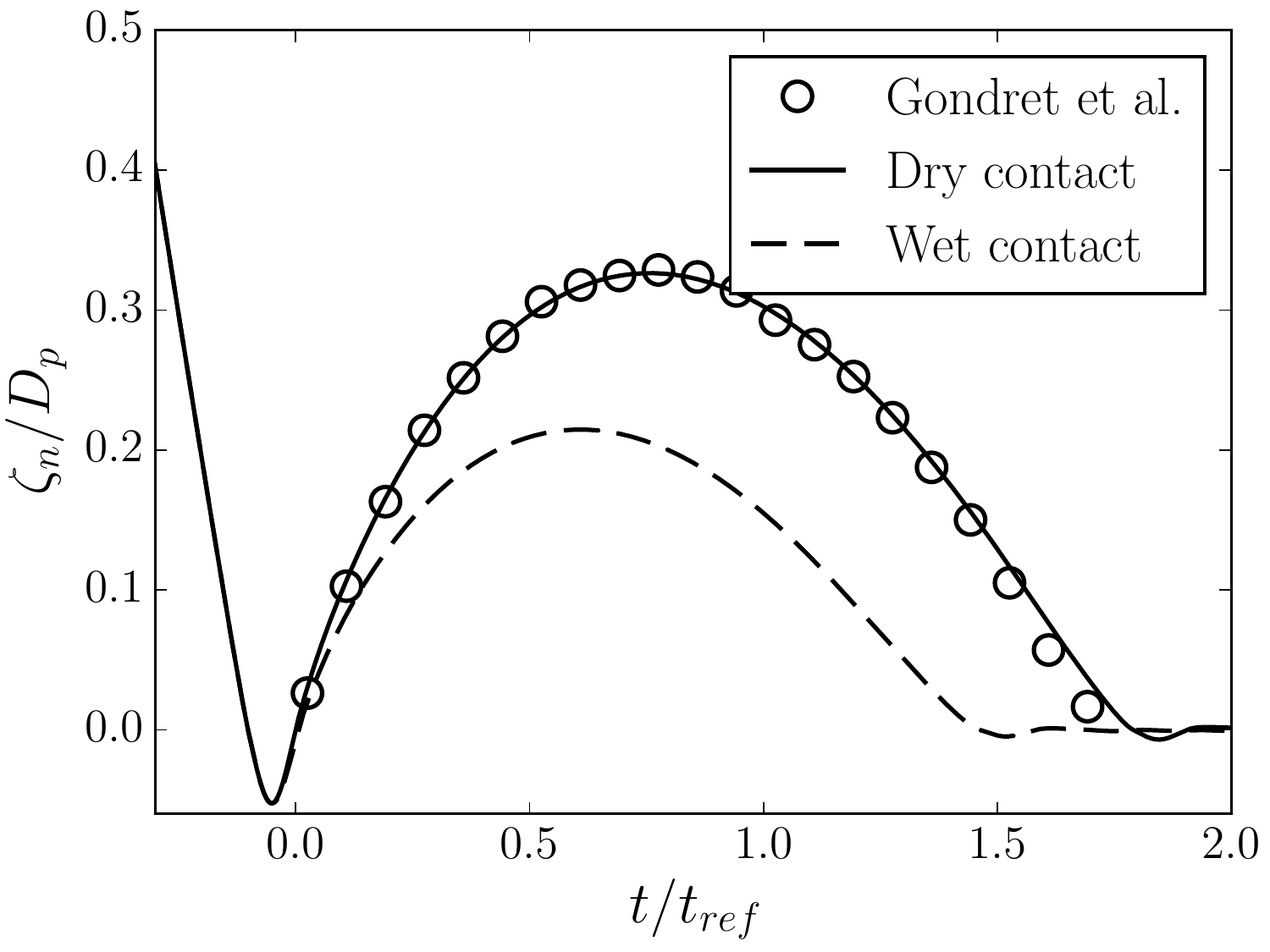}}
    \put( 0.0,  4.2)     {a) }
    \put( 6.5,  4.2)     {b) }
\end{picture}
        \caption{\small \textit{Effect of including (``wet") or excluding (``dry") fluid forces during contact on rebound trajectories. a) $\mrm{St} = 152$ and b) $\mrm{St} = 27$.}}
    \label{fig:gondret_wet_dry}
\end{figure}
On the other hand, including the fluid forces during contact can lead to significant drag on the particle throughout the collision.  Figure~\ref{fig:gondret_wet_dry} shows how excluding fluid forces during contact gives us excellent agreement with the experimental results, while including fluid forces during contact leads to excessive damping because the fluid surrounding the particle has not been able to adapt to the change of the kinematics of the particle.
Indeed, the simulations of \cite{simeonov2012} show coefficients of restitution below experimental values for moderate Stokes numbers ($20 < \mrm{St} < 100$).
Hence, we decided to follow \cite{kempe2012a} and to exclude fluid forces during contact for collisions with $\mrm{St} \gg 1$, redefining \eqref{eq:part_lin} as follows:
\begin{linenomath*}
\begin{equation} \label{eq:p_eom_acm}
m_p\: \frac{\text{d}\textbf{u}_p}{\text{d} t} = \begin{cases}
	\mbf{F}_{h,p} + \mbf{F}_{g,p} + \mbf{F}_{c,p} & \zeta_n > 0 \\
	\mbf{F}_{g,p} + \mbf{F}_{c,p} & \zeta_n \leq 0 \qquad .
	\end{cases}
\end{equation}
\end{linenomath*}
\cite{costa2015} implemented a similar method for particle-wall collisions, but they turned off fluid forces when the collision overlap exceeded the expected overlap due to the particle's weight, i.e. $\zeta_n < -(1 - \rho_f/\rho_p) g m_p / k_n$.
For the cases shown in Figure~\ref{fig:gondret_wet_dry}, the timescale of the contact phase is much smaller than the timescale of the general fluid flow, i.e. the timescale of the particle rebound.  Thus, while neglecting fluid forces has an important effect on realizing the correct $e_{wet}$, it has a minimal effect on the general flow.

However, neglecting fluid forces can lead to unphysical situations for enduring contact, which we define to be when the timescale of contact matches or exceeds that of the general flow.  Consider, for example, a single particle at rest and in contact with a wall.  If we then impose a shear flow over the particle, it should be swept up into the flow, or at the very least be carried downstream.  However, in a simulation using \eqref{eq:p_eom_acm}, because the particle is in contact with the wall, it does not experience the hydrodynamic forces.  It will therefore continue to sit on the wall, oblivious to the flow around it, until another particle collides with it.
This was addressed in \cite{kempe2014} by switching on the hydrodynamic forces for all collisions regardless of the Stokes number, even though it was not explicitly mentioned in this reference (Kempe \& Fr\"ohlich, 2016, private communication).
We address this problem in detail in the subsequent Section \ref{sec:enduring_contact} to introduce a suitable threshold for the inclusion of the hydrodynamic forces in \eqref{eq:part_lin} and \eqref{eq:part_ang}.

\section{Enduring contact model}\label{sec:enduring_contact}

\subsection{Accounting for fluid forces}

As shown in the results from Section \ref{sec:enhancements}, neglecting fluid forces acting on the particle during contact produces a good match with the experimental data of \cite{gondret2002}, which involve collisions of finite duration.  However, problems can arise in the limit of enduring contact.  We therefore propose to include fluid forces during contact below some threshold Stokes number $\mrm{St}_{crit}$.  For collisions above $\mrm{St}_{crit}$, the contact duration should be finite ($T_c = 10\Delta t$) so that no major loss of physicality is encountered.  For collisions below $\mrm{St}_{crit}$, the particle is not going to experience an appreciable rebound so that the particle motion is not governed by collision forces during contact, but by hydrodynamic forces. Neglecting hydrodynamic forces in the low-Stokes number regime introduces artifacts in particle mobility.
Indeed, this was observed in \cite{vowinckel2016} for the situation of a horizontal turbulent open-channel flow laden with particles heavier than their critical threshold of motion. Using the same method for collisions, these particles formed a closed bed of resting particles. In this reference, it was shown that a collision with a fast moving particle was necessary for almost all of the erosion events recorded to dislodge a particle out of the sediment packing.  However, it has not been possible to clarify to what extent this triggering collision is merely a consequence of the collision procedure.

To investigate what the critical value for the Stokes number may be, we compared particle-wall collisions that include hydrodynamic forces during contact (``wet" contact) to those that exclude hydrodynamic forces during contact (``dry" contact), as illustrated in Figure~\ref{fig:wetdry}.  For this scenario, we used the same parameters as those summarized in Table \ref{tab:gondret} ($\mrm{St}=27$) and repeated the simulations for ever-decreasing $\mrm{St}$.  The Stokes number was controlled by prescribing the particle's velocity until it made direct contact with the wall.  Unlike the previous simulations, we did not allow the lubrication layer to slow the particle before contact.

\setlength{\unitlength}{1cm}
\begin{figure}[t]
\begin{picture}(7,8.6)

  \put(0,   4.4 ){\includegraphics[width=0.50\textwidth]{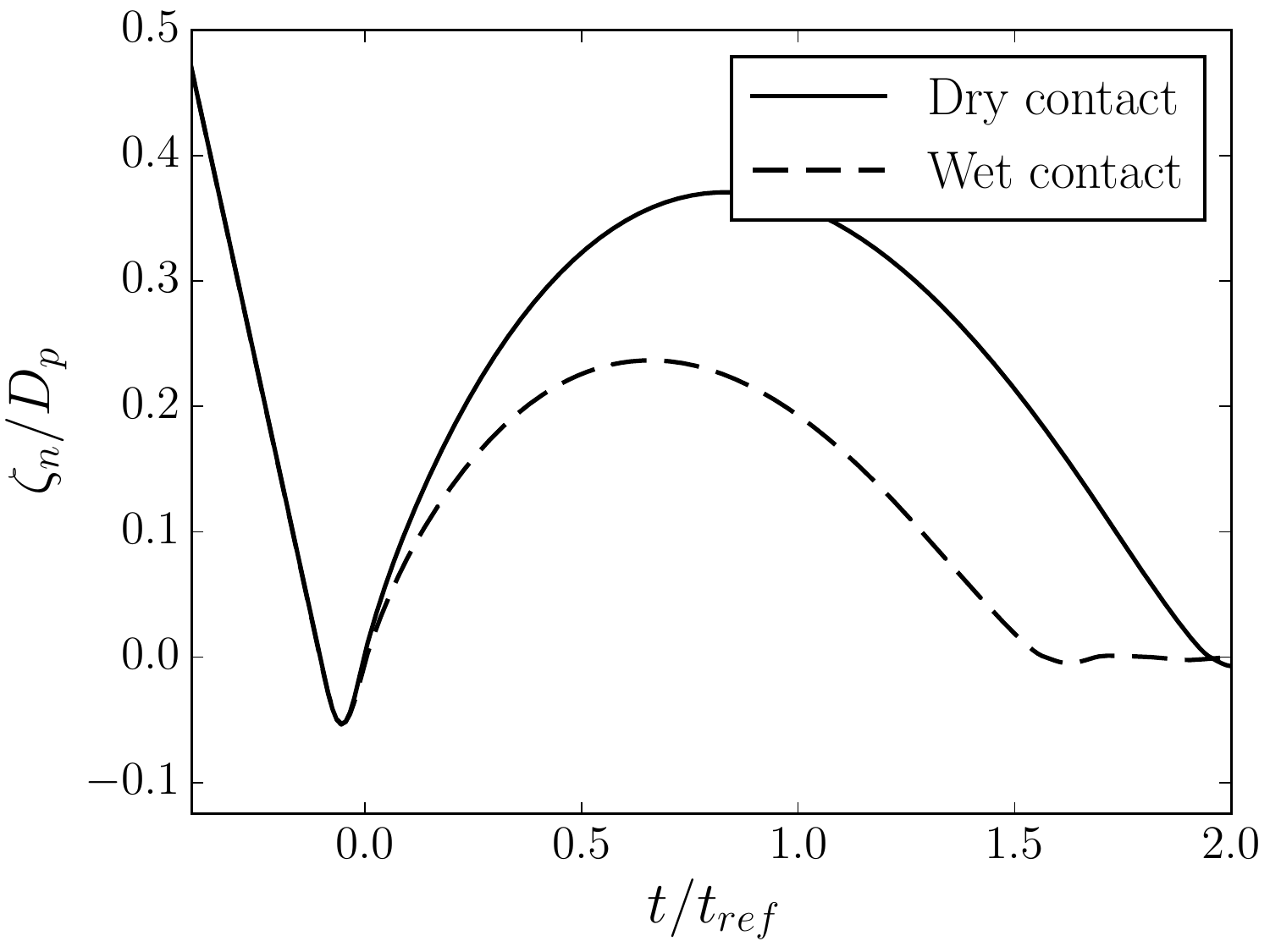}}
  \put(6.5, 4.4 ){\includegraphics[width=0.50\textwidth]{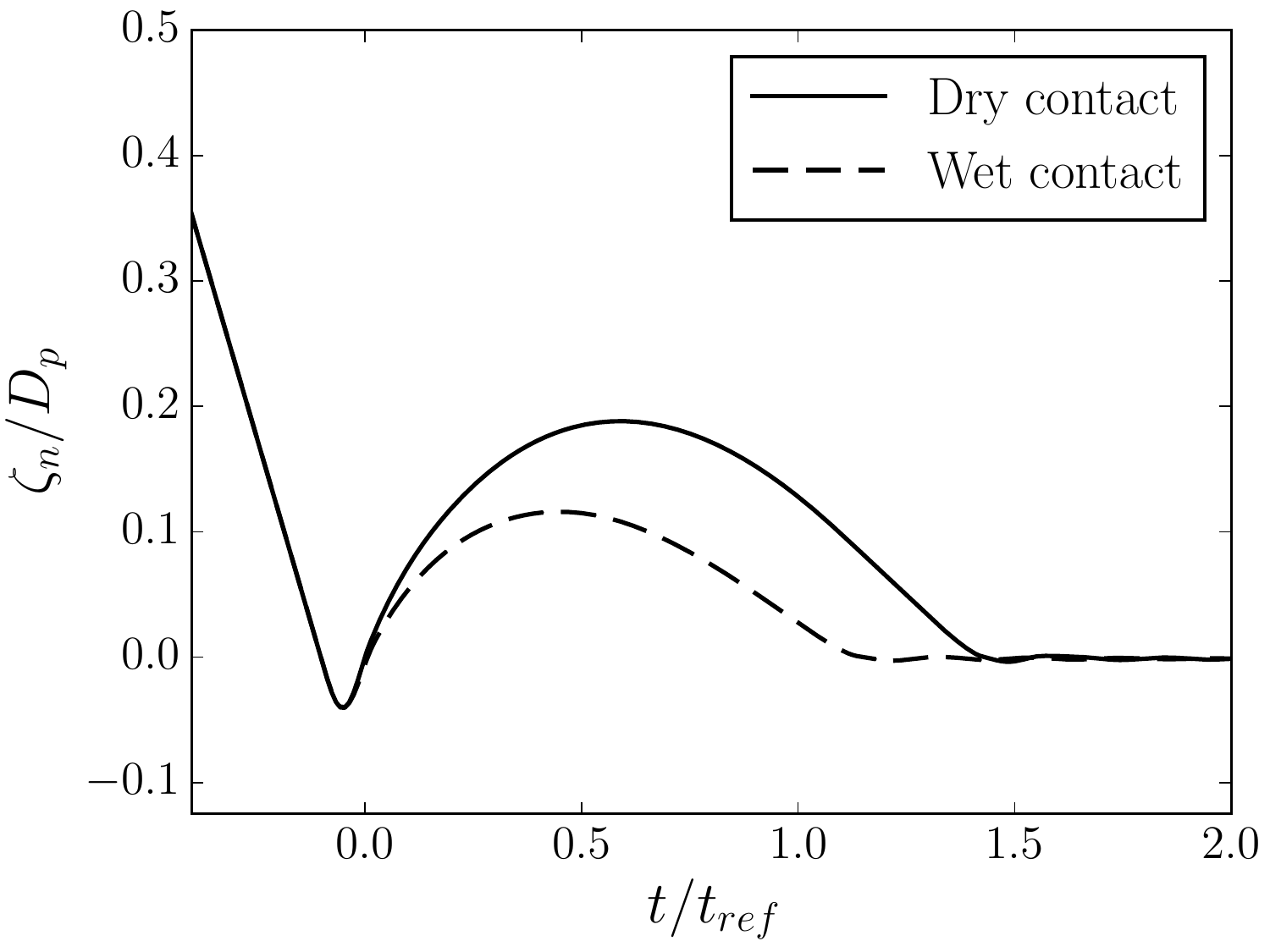}}
  \put(0,  -0.2 ){\includegraphics[width=0.50\textwidth]{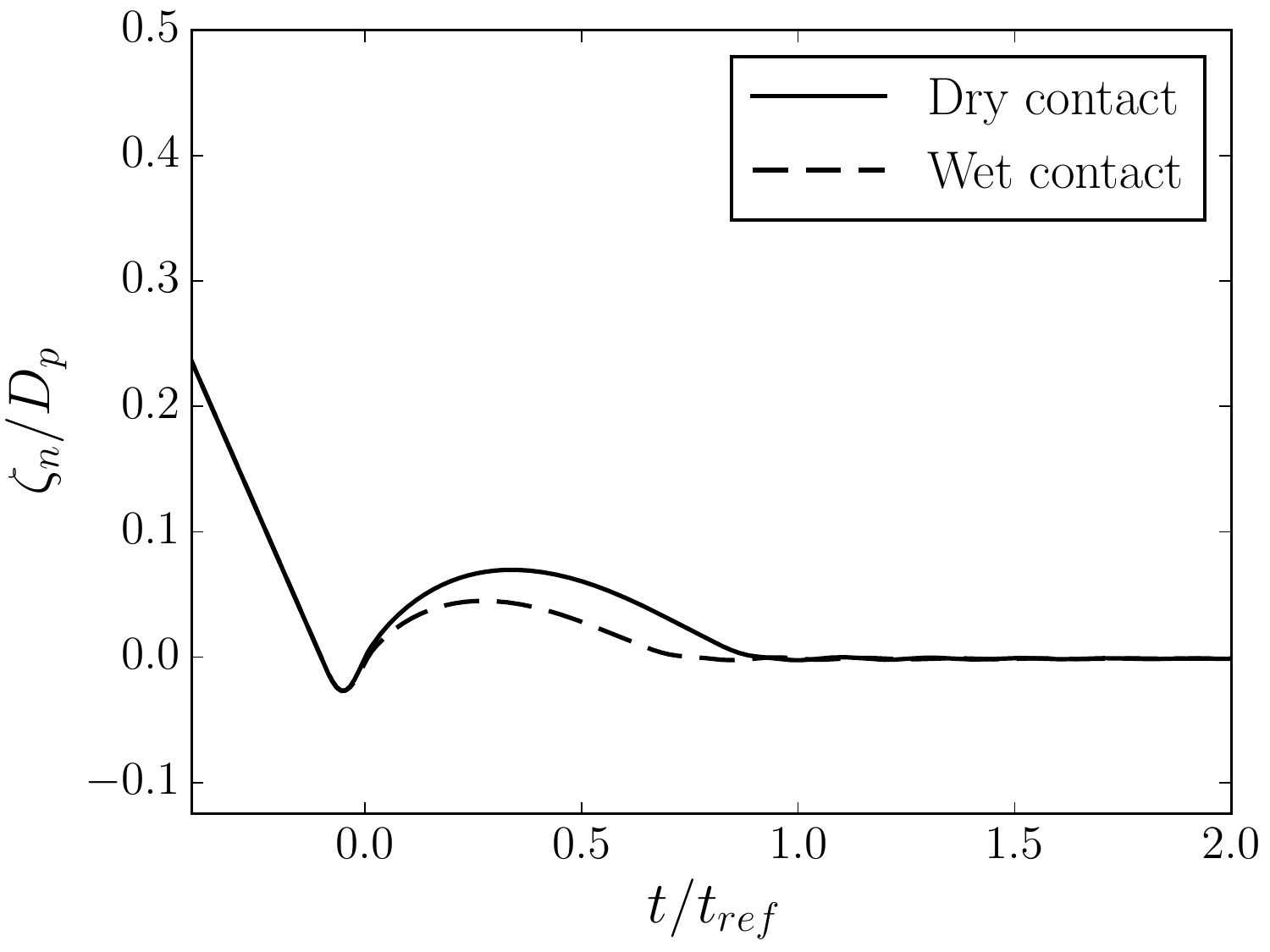}}
  \put(6.5,-0.15){\includegraphics[width=0.50\textwidth]{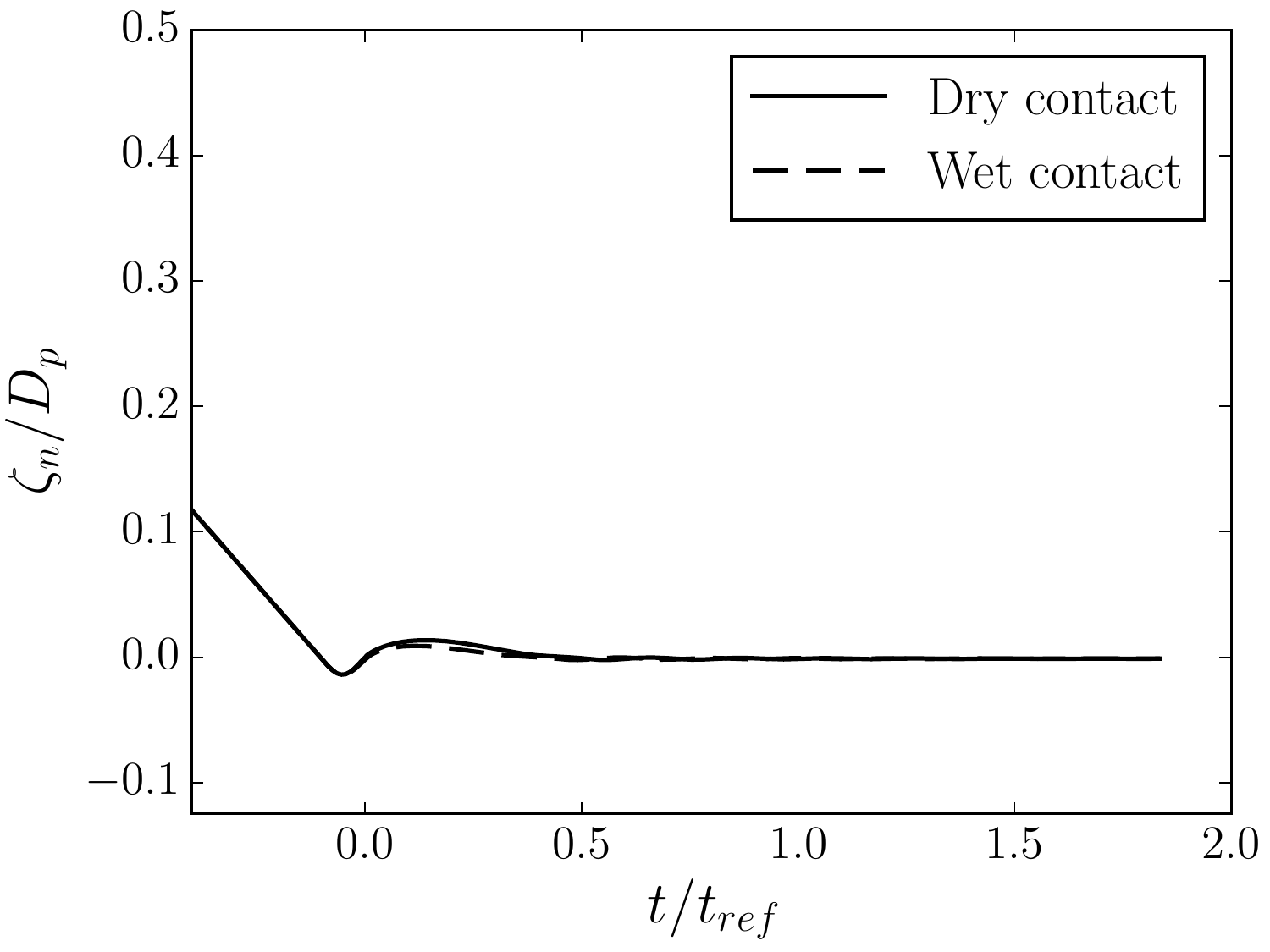}}
    \put( 0.0,  8.6)     {a) }
    \put( 6.5,  8.6)     {b) }
    \put( 0.0,  4.0)     {c) }
    \put( 6.5,  4.0)     {d) }

\end{picture}
        \caption{\small \textit{Comparison of trajectories of particle-wall collisions without (``dry'') and with (``wet'') hydrodynamic forces during contact for various Stokes numbers a) $St = 20$, b) $St = 15$, c) $St = 10$, d) $St = 5$, where $u_{in}$ is measured at $\zeta_n = 0$.}}
    \label{fig:wetdry}
\end{figure}
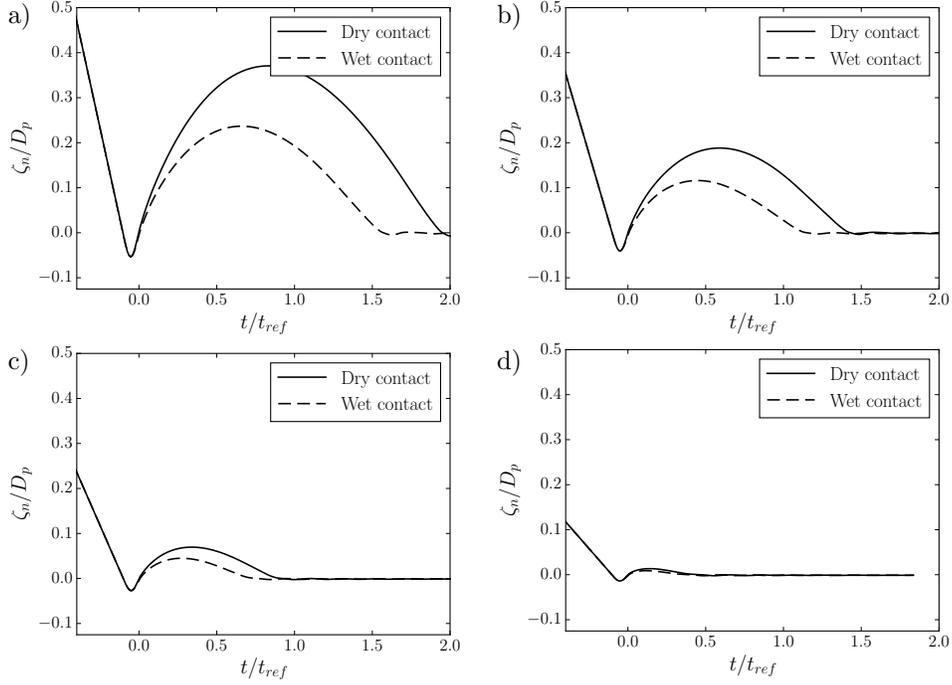

For the cases with higher Stokes numbers, we can clearly see how including hydrodynamic forces during contact leads to significant undershooting of the rebound trajectory.  As the Stokes number decreases, however, the significance of this undershooting also decreases.  For $\mrm{St} < 5$ (Figure~\ref{fig:wetdry}d), there is no appreciable rebound, and we consider the particle to be in enduring contact.  Thus, based on these plots, we selected the critical Stokes number to be $\mrm{St}_{crit} = 5$.  This value is consistent with the work of other researchers \citep{gondret2002, joseph2001}, who experimentally observed no rebounds for $\mrm{St} < 10$.
Note that the Stokes numbers reported in Figure~\ref{fig:wetdry} and our resulting $\mrm{St}_{crit}$ are based on the particle velocity at contact, i.e. when $\zeta_n = 0$, whereas most other authors report Stokes numbers at some distance from the wall, before the lubrication layer has fully slowed the particle.
With this enduring contact model, we can expand the particle equation of motion \eqref{eq:p_eom_acm} to
\begin{linenomath*}
\begin{equation} \label{eq:eom_cases}
m_p\: \frac{\text{d}\textbf{u}_p}{\text{d} t} = \begin{cases}
	\mbf{F}_{h,p} + \mbf{F}_{g,p} + \mbf{F}_{c,p} , &  \zeta_n > 0 \\
	\mbf{F}_{g,p} + \mbf{F}_{c,p} , & \zeta_n \leq 0 \wedge \max\{\mrm{St}\}>\mrm{St}_{crit} \\
	\mbf{F}_{h,p} + \mbf{F}_{g,p} + \mbf{F}_{c,p} , & \zeta_n \leq 0 \wedge \max\{\mrm{St}\}\leq\mrm{St}_{crit} \qquad,
	\end{cases}
\end{equation}
\end{linenomath*}
where the $\max\{\mrm{St}\}$ function represents the maximum Stokes number among all active collisions for particle $p$, and fluid forces acting on the particle are only included from non-disabled markers. The same consideration applies for the angular momentum \eqref{eq:part_ang}. Using this scheme now allows us to include the full momentum balance for particles in enduring contact, i.e. the hydrodynamic stresses as well as the buoyant weight of the particle, so that the considerations of the Shields parameter become applicable.

\subsection{Optimizing enduring particle overlap}\label{sec:overlap}

In the case of $\mrm{St} \ll 1$, the impact velocity $u_{in}$ approaches zero. This means in turn that the computed stiffness in \eqref{eq:k_n_explicit} would approach infinity.  This problem is addressed by \cite{kempe2012a} who have introduced a critical Stokes number $\mrm{St}_{crit}$, which establishes a minimum impact velocity to limit $k_n$ for enduring contact:
\begin{linenomath*}
\begin{equation}
u_{in,crit} = \frac{ 9 \mrm{St}_{crit} \, \rho_f \nu_f }{ \rho_p D_p } \qquad.
\end{equation}
\end{linenomath*}
In the present study, this critical impact velocity was used in \eqref{eq:k_n_explicit} and \eqref{eq:d_n_explicit} to compute $k_n$ and $d_n$, respectively, for such collisions. This implementation differs slightly from that of \cite{kempe2012a}, who do not apply any damping for collisions with $\mrm{St}<\mrm{St}_{crit}$, i.e. they have set $d_n=0$.  We included this damping for enduring contact in order to help reach steady-state conditions.  Our implementation also differs in that we use $\mrm{St}_{crit} = 5$ whereas \cite{kempe2012a} used $\mrm{St}_{crit} = 1$.

Furthermore, we retain the buoyant weight forces in the equation of motion during contact as outlined in Section \ref{sec:neglecting_fluid_force}.  This means that, for particle packings several diameters thick, the weight of a single sphere resting on another layer of particles is passed along to deeper layers. This effect enhances the physical realism because frictional contact forces increase with depth, but it also results in increasing particle surface overlap with depth and ultimately in a change of porosity of the sediment bed, which has been acknowledged as a crucial parameter to define the hydraulic resistance of a sediment to the flow \citep{vowinckel2014}.
However, a flow with a lower Reynolds number would result in collisions with lower Stokes numbers such that $u_{in,crit}$ could become large relative to the particle size and relevant time scales.  A large $u_{in,crit}$ would result in a low $k_n$ and hence a large overlap between particles, which is undesirable.  To prevent this large overlap, we enforce a maximum overlap distance $\epsilon R_p$ through the following procedure: for a collision with $\mrm{St}<\mrm{St}_{crit}$, the stiffness is given by
\begin{linenomath*}
\begin{equation}
k_n = \begin{cases}
	\frac{m_\textit{eff}}{\sqrt{u_{in}t_*^5}} & u_{in} > u_{in,crit} \\
	\max \left( k_{n,crit}, \, k_{n,grav} \right) & u_{in} < u_{in,crit}
	\end{cases}
\end{equation}
\end{linenomath*}
where $m_\textit{eff}$ and $t_*$ are defined in \eqref{eq:m_eff} and \eqref{eq:t_explicit}, respectively,
\begin{linenomath*}
\begin{equation}
k_{n,crit} = \frac{m_\textit{eff}}{\sqrt{u_{in,crit}t_*^5}}
\end{equation}
\end{linenomath*}
is the stiffness limited by the critical impact velocity, and
\begin{linenomath*}
\begin{equation}
k_{n,grav} = \max \left[ m_p g ( \epsilon R_p )^{-3/2}, \, m_q g ( \epsilon R_q )^{-3/2} \right]
\end{equation}
\end{linenomath*}
is the stiffness required for particle $p$ (or $q$) to have a steady-state overlap of $\epsilon R_p$ (or $\epsilon R_q$) with a wall due to gravity. To have a minimal constant overlap we set $\epsilon = 10^{-3}$.  Thus, we ensure that a bed of particles contains a uniform set of collision stiffnesses that minimize particle overlap.

\subsection{Rolling and sliding motion}\label{sec:rolling_sliding}
%
The coefficient of friction for a material can depend on whether the contact is rolling or sliding \citep{fishbane1996}. The rolling condition implies zero slip at the contact point, i.e. $\lVert \textbf{g}_{t,cp}\rVert=0$ (cf. \ref{app:definitions}). As a consequence, particle surfaces are in sticking contact for rolling motion until a critical threshold of static friction $F_s= \mu_s \lVert  \textbf{F}_{n} \rVert$ is exceeded, where $\mu_s$ is  the coefficient of static friction. As soon as this condition is met, significant slip occurs and the contact condition changes from sticking to sliding, so that the threshold for kinetic friction $F_k= \mu_k \lVert  \textbf{F}_{n} \rVert$ must be used, where $\mu_k$ is the coefficient of kinetic friction, with $\mu_s$ always greater than $\mu_k$. Apart from the physical reasoning presented above, limiting the frictional forces also becomes important from a numerical point of view whenever two or more collision partners are involved.  Otherwise the multiple contact points competing for no-slip conditions can lead to instabilities in the calculation of the frictional forces.

%
%
\begin{table}[t]
\centering
\begin{tabular}{l c c}
\hline
\hline
Case & Dry oblique collision & Rolling in shear flow \\
\hline
$R_p$				& 0.00159	& 0.0625	\\
$\rho_p / \rho_f$	& 2500		& 2.5		\\
$e_{dry}$			& 0.83		& 0.97		\\
$\nu$				& 0.22		& 0.3		\\
$\mu_k$				& 0.11		& 0.15		\\
$\mu_s$				& 0.8		& 0.8		\\
$g$					& 0			& 9.81		\\
$\nu_f$				& 0			& 0.02		\\
\hline
Timestep	& $\Delta t = 2\mrm{e-}5$	& CFL = 0.5		\\
\hline
\hline
\end{tabular}
\caption{\small \textit{Simulation setup for oblique and rolling sphere simulations.
}}
\label{tab:tangential_sim}
\end{table}

In the present study, the distinction between rolling/sticking and sliding is made by the following scheme, which is comparable to that of \cite{luding2008}:
\begin{itemize}
\item While the particle is sticking, i.e. $||\mbf{F}_{t,LS}|| < ||\mu \mbf{F}_n||$, we set $\mu = \mu_s$ to test for the onset of slipping.

\item Once slipping occurs, i.e. $||\mbf{F}_{t,LS}|| > ||\mu \mbf{F}_n||$, we set $\mu = \mu_k$ until the friction force falls below the Coulomb friction force.
\end{itemize}
The aim of the present study is to simulate natural sediment. Hence we parametrized the coefficients of friction with typical values of silicate materials, yielding $\mu_k = 0.15$ based on the work of \cite{joseph2004}, who worked with glass spheres, and $\mu_s = 0.8$ based on the work of \cite{dieterich1972}, who found values ranging from 0.75 to 0.85 for different rock materials like quartz, granite, and sandstone.

\begin{figure}[t]
\centering
\includegraphics[width=0.65\textwidth]{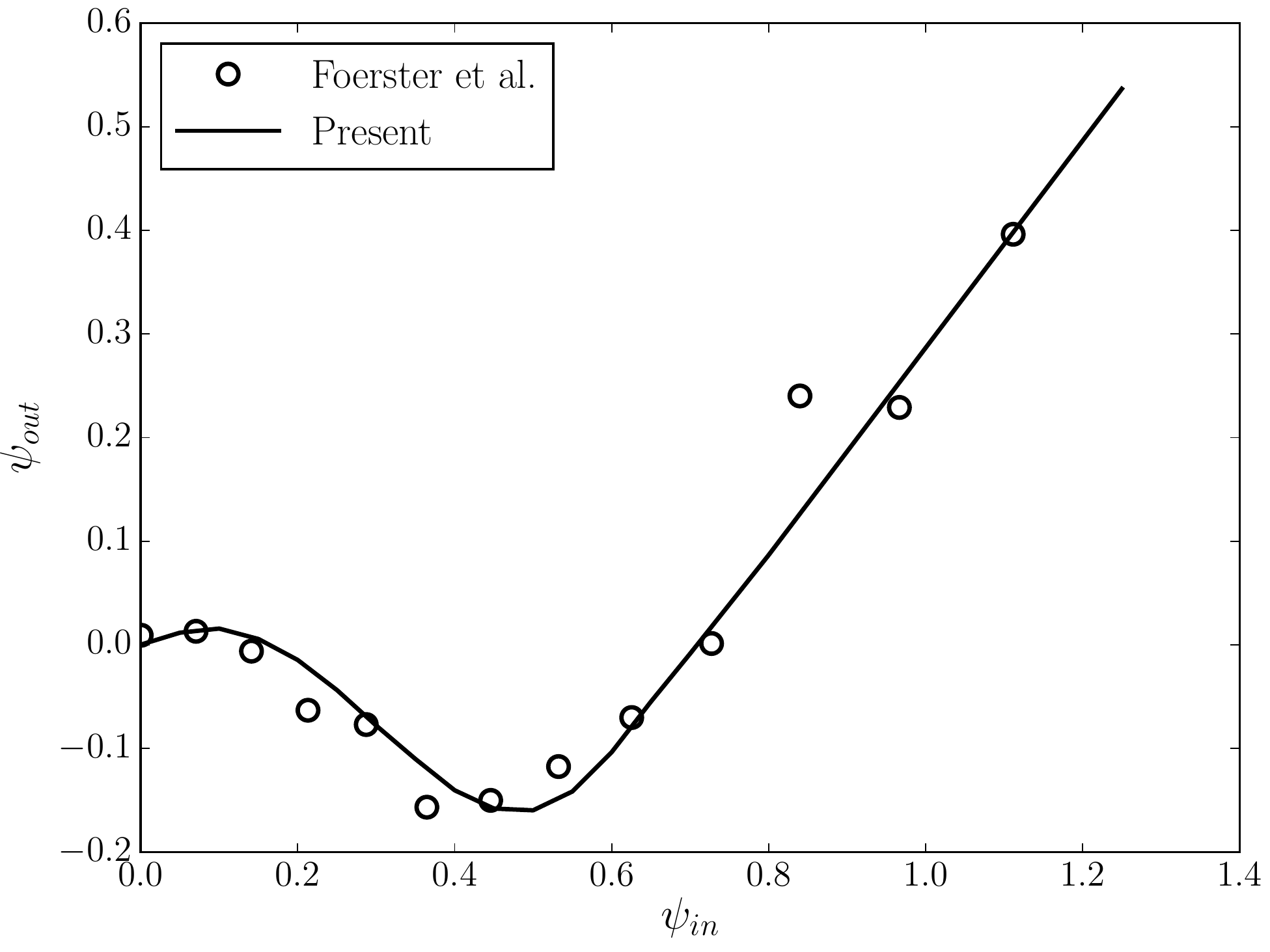}
\caption{\small \textit{Rebound vs. impact angles for a particle-wall oblique collision.}}
\label{fig:oblique_impact}
\end{figure}

We have validated the tangential collision model using an oblique dry impact experiment, i.e. neglecting hydrodynamic forces, by \cite{foerster1994}, whose parameters are summarized in Table~\ref{tab:tangential_sim}.  Figure~\ref{fig:oblique_impact} shows that our simulations compare well to the experiments in reproducing the rebound angle
\begin{linenomath*}
\begin{equation} \label{eq:rebound_angle}
\psi_{out} = \frac{u_{t,out}}{u_{n,in}} \qquad ,
\end{equation}
\end{linenomath*}
which depends on the impact angle
\begin{linenomath*}
\begin{equation} \label{eq:impact_angle}
\psi_{in} = \frac{u_{t,in}}{u_{n,in}} \qquad .
\end{equation}
\end{linenomath*}
Here, $u_{n,in}$ is the impact velocity normal to the wall, while $u_{t,in}$ and $u_{t,out}$ are the impact and rebound velocities, respectively, of the particle's contact point tangential to the wall ($u_t = u_p + R_p \, \omega_{p,z}$ for a particle obliquely colliding in the $x$-direction).
For a particle with no initial rotation, $\psi_{in}$ is the tangent of the angle the particle makes with the wall from the normal ($\psi_{in} = 0$ means no relative tangent motion).  The rebound angle is zero when the contact is sticking perfectly at the time of release.  However, the rebound angle is negative when, at the point of release, $|u_p| < |\omega_{p,z}|$ (since $\omega_{p,z} < 0$ for our example).  The linear-spring tangential collision model is able to perfectly capture these negative values for $\psi_{out}$ at low impact angles.

%
To test both situations, rolling and sliding, we simulated a particle in a Couette flow. We placed a sphere of radius $R_p/H = 0.0625$ on the bottom wall of a channel of height $H$. We initialized the particle at rest at a distance $\zeta_n/R_p = 1.6 \times 10^{-5}$ above the bottom wall.  We subsequently exposed the sphere to a linear shear flow, holding it fixed for a short time ($t U / H = 0.01$) to allow the flow to develop around it before releasing it. The numerical parameters are summarized in Table~\ref{tab:tangential_sim}.  We found that slipping motion occurs for a lower Reynolds number of $\mrm{Re}_H = U H / \nu_f= 10$, where $U$ is the lid velocity.  On the other hand, perfect rolling motion occurs at $\mrm{Re}_H = U H / \nu_f= 50$.

%
%
\setlength{\unitlength}{1cm}
\begin{figure}[t]
\begin{picture}(7,4.5)
  \put(0,  -0){\includegraphics[width=0.5\textwidth]{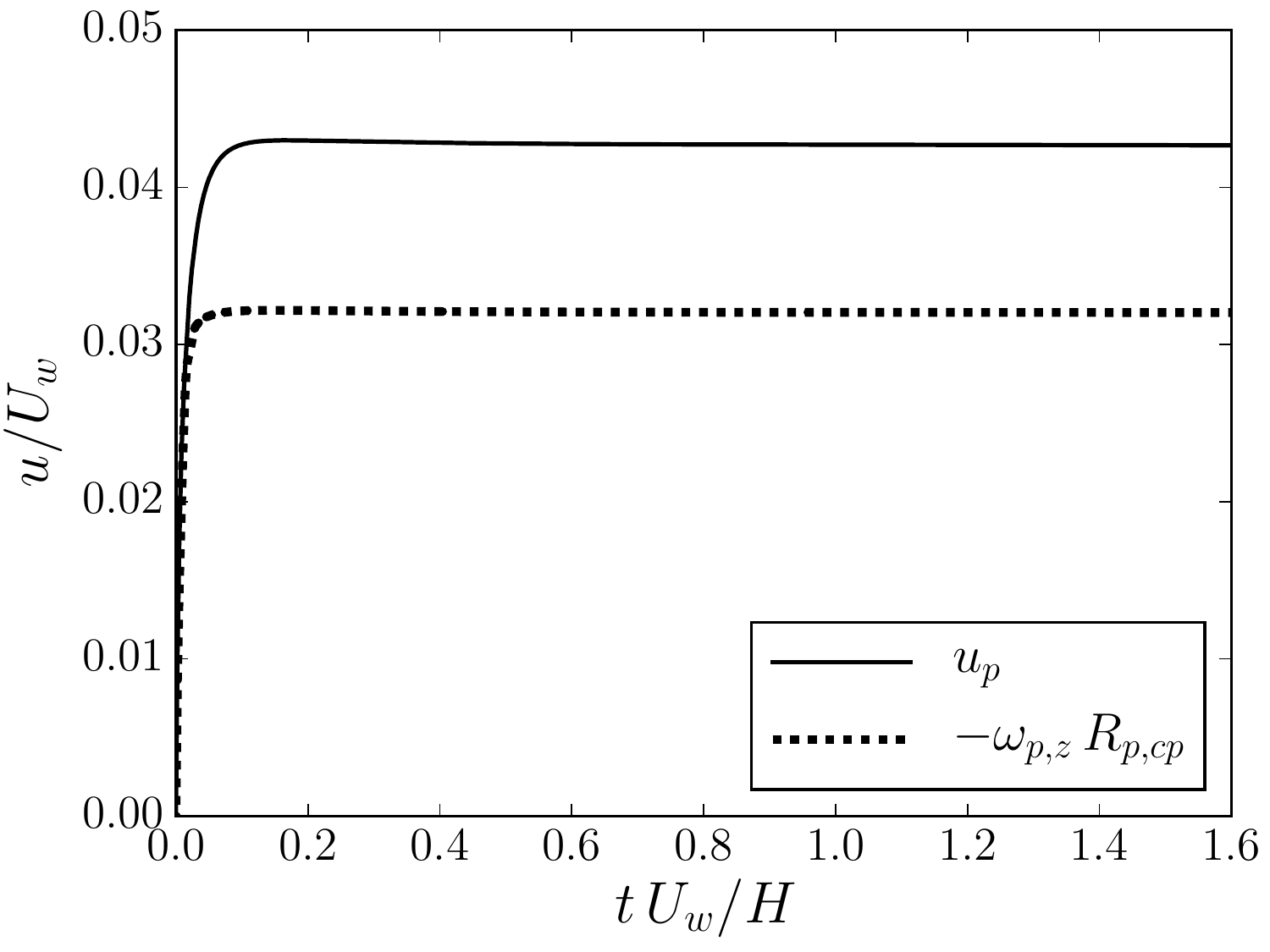}}
  \put(6.5,-0){\includegraphics[width=0.5\textwidth]{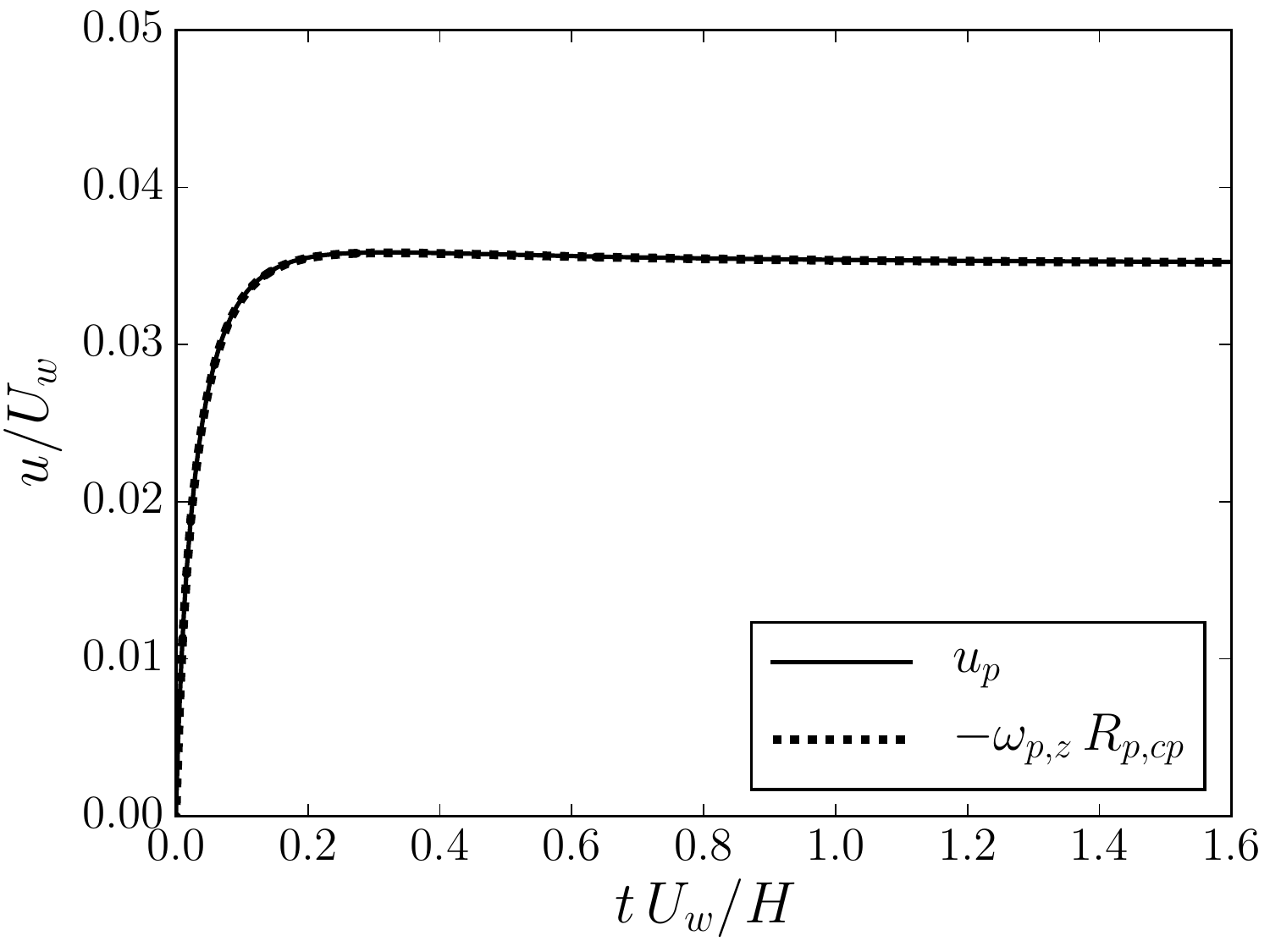}}
    \put( 0.0,  4.2)     {a) }
    \put( 6.5,  4.2)     {b) }
\end{picture}
        \caption{\small \textit{Translational and rotational velocities of a particle exposed to a linear shear flow. a) sliding motion for $Re_H = 10$ and b) rolling motion for $Re_H= 50$.}}
    \label{fig:rolling_sphere}
\end{figure}
Figure~\ref{fig:rolling_sphere} shows how the particle accelerates until it reaches a steady-state translational velocity. As soon as the particle makes contact with the wall, gravity holds it there with a slight overlap according to the conditions defined in Section \ref{sec:overlap}.  Accounting for fluid forces during contact allows the particle to accelerate to a steady-state speed while in contact with the wall. As expected the particle achieves perfect rolling without slip (Figure~\ref{fig:rolling_sphere}b), marked by the match between the translational velocity $u_p$ and the rotational velocity relative to the particle center $-\omega_{p,z} \, R_{p,cp}$.  Accordingly, the particle shows significant slip for the lower Reynolds number (Figure~\ref{fig:rolling_sphere}a), where the increased viscosity leads to increased drag on the particle, which in turn overpowers the friction from the particle's weight.

\section{Flow over dense sediment}\label{sec:bulk_flow}

\subsection{Physical setup}

%
\begin{figure}[t]
\centering
\includegraphics[width=0.95\textwidth]{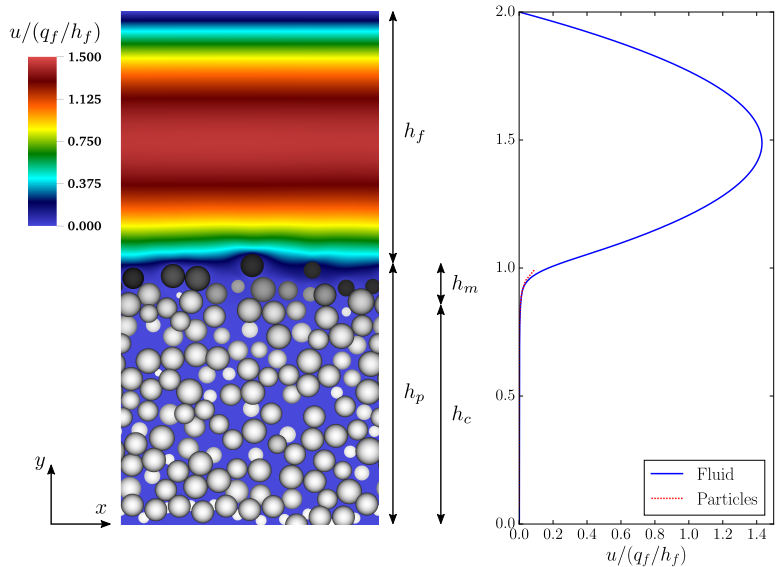}
\caption{\small \textit{Left plot: Instantaneous snapshot of a slice through the $xy$-plane for case A10M. Contours show the streamwise component of the fluid velocity.  Particles are colored grayscale according to their velocity.  Right plot: Streamwise and spanwise averages of fluid and particle velocities.  Arrows correspond to the length scales for the clear fluid, $h_f$, the particle bed, $h_p$, the mobile bed layer, $h_m$, and the motionless bed layer, $h_c$.}}
\label{fig:aussillous1}
\end{figure}

%
\begin{table}[t]
\centering
\begin{tabular}{lcc}
\hline
\hline
$\mrm{Ga}$			& 0.397  \\
\hline
$R_p$ (m)			& 0.0444  \\
$\rho_p / \rho_f$	& 2.1  \\
$\nu_f$ (m$^2$/s)		& 0.219  \\
$g$ (m/s$^2$)		& 9.81  \\
$e_{dry}$			& 0.97  \\
$\zeta_{n,min}$ (m)	& $3.0 \times 10^{-3} R_p$  \\
$\mu_k$				& 0.15  \\
$\mu_s$				& 0.8  \\
\hline
Domain size (m) ($L_x \times L_y \times L_z$)
			& $1.0 \times 2.0 \times 1.0$  \\
Domain boundary conditions	& $\mrm{p} \times \mrm{ns} \times \mrm{p}$  \\
Grid cells in $x$-direction	& 256  \\
$D_p / h$		& 22.7  \\
Volume fraction in center of bed	& 0.609  \\
Timestep		& $\mrm{CFL} = 0.1$  \\
\hline
\hline
\end{tabular}
\caption{\small \textit{Simulation parameters to match the experiments of \cite{aussillous2013}.  Boundary conditions are periodic (p) and no-slip (ns).  The Galileo number $\mrm{Ga}$ is defined in~\eqref{eq:Ga}.
}}
\label{tab:aussillous}
\end{table}
%

%
\begin{table}[t]
\centering
\begin{tabular}{lccccc}
\hline
\hline
Case			& A1		& A2		& A9		& A10 		& A10M\\
$\mrm{Re}_b$		& 0.301		& 0.402		& 1.01	& 1.15	& 1.15 \\
$h_f / D_p$ (Exp.)	& $7.05 \pm 0.5$	& $8.15 \pm 0.5$	& $10.29 \pm 0.5$	& $11.27 \pm 0.5$
	& $11.27 \pm 0.5$ \\
$h_f / D_p$ (Sim.)	& $7.15$			& $8.31$			& $10.33$			& $11.29$
	& 11.05 \\
$\mrm{Sh}$ (Exp.)	& $0.24 \pm 0.03$	& $0.24 \pm 0.03$	& $0.37 \pm 0.04$	& $0.35 \pm 0.03$
	& $0.35 \pm 0.03$ \\
$\mrm{Sh}$ (Sim.)	& $0.224$			& $0.222$			& $0.358$			& $0.343$
	& 0.357 \\
\hline
$q_f$ (m$^2$/s)		& 0.0659	& 0.0880	& 0.220	& 0.251		& 0.251 \\
$N_{p,m}$		& 2031	& 1870	& 1559	& 1419	& $1407^*$ \\
$N_{p,f}$		& 132	& 132	& 132	& 132	& 132 \\
$T_\textit{avg}$ (s)	& 139.5	& 137.9	& 126.2	& 127.3 & 111.0 \\
\hline
\hline
\end{tabular}
\caption{\small \textit{Parameters that vary between the different cases.  The bulk Reynolds number $\mrm{Re}_b$ is defined in \eqref{eq:Reb} and the Shields number $\mrm{Sh}$ is defined in \eqref{eq:Sh}.  The fluid height $h_f$ (and hence Shields number) do not exactly match between the experiments (Exp.) and simulations (Sim.). $^*$Polydisperse particle diameters follow a Gaussian distribution with a standard deviation of $\sigma(D_p) = 0.1 D_p$.
}}
\label{tab:aussillous_flow_rate}
\end{table}

We presented a detailed validation of binary particle-wall collisions in Sections \ref{sec:enhancements} and \ref{sec:enduring_contact}. To address the bulk behavior of a dense granular bed sheared by a laminar Poiseuille flow, we carried out numerical simulations to reproduce the experimental results of \cite{aussillous2013}, who studied pressure-driven flows over glass spheres with a mean diameter $D_p = 1.1$mm and a standard deviation of $\sigma(D_p) = 0.1$mm as sediment material.
This experimental work provides investigations over a range of submergences $h_f/D_p$ and Reynolds numbers in the laminar regime, where $h_f$ is the height of the clear-water layer above the sediment bed illustrated in Figure~\ref{fig:aussillous1}.
We define $h_f$ to be the height above which the average particle volume fraction $\phi<0.05$, which is the threshold for negligible impact of particle-particle interaction on the flow \citep{capart2011}.  We define the mobile bed height $h_m$ to be the portion of the particle bed above which the mean particle velocity is higher than 1\% of the value at the fluid/particle interface.

In their experiments, \cite{aussillous2013} filled a long chamber with particles and then applied a constant pressure gradient, which eroded the particles from the chamber.  Initially, the fluid height $h_f$ was small and the pressure gradient drove a large number of particles so that the height of the mobile bed layer, $h_m$ in Figure~\ref{fig:aussillous1}, was large.  Since no new particles were added to the chamber, $h_f$ increased as the particles eroded away until, at long periods of time, the experiment reached a steady-state configuration where the influx of particles into the observation window remained in equilibrium with the outflux.  Due to our use of periodic boundary conditions, we will only try to replicate the long-term steady-state flow conditions, of which there are only a few data from \cite{aussillous2013}

We executed several simulations in an attempt to match four of the experiments of \cite{aussillous2013} at different flow rates and fluid heights.
The physical and numerical parameters associated with these simulations are listed in Table~\ref{tab:aussillous}, and the differences between the four cases are listed in Table~\ref{tab:aussillous_flow_rate}.  These experiments can be characterized by the Galileo number
\begin{linenomath*}
\begin{equation} \label{eq:Ga}
\mrm{Ga} = \frac{\sqrt{(\rho_p/\rho_f - 1) g D_p^3}}{\nu_f} \qquad,
\end{equation}
\end{linenomath*}
the bulk Reynolds number
\begin{linenomath*}
\begin{equation} \label{eq:Reb}
\mrm{Re}_b = \frac{q_f}{\nu_f} \qquad,
\end{equation}
\end{linenomath*}
where $q_f$ is the fluid flow rate, and the Shields number
\begin{linenomath*}
\begin{equation} \label{eq:Sh}
\mrm{Sh} = \frac{6 \mrm{Re}_b}{\mrm{Ga}^2} \left( \frac{D_p}{h_f} \right)^2 \qquad,
\end{equation}
\end{linenomath*}
which represents the ratio of the shear stress acting on the particle bed to the buoyant weight of a particle.  \cite{aussillous2013} reported an uncertainty for the determination of the bed height as $h_f \pm R_p$, which we have included in Table~\ref{tab:aussillous_flow_rate} as the deviations in $h_f$ and $\mrm{Sh}$, which depends on $h_f$.

We required a low $\mrm{CFL} = 0.1$ in order to maintain the stability of the fluid-particle coupling.  This restricted CFL value was necessary to avoid numerical instabilities arising from the simultaneous particle-particle interactions of a multitude of particles within the thick sediment bed.
We also used a grid resolution of $D_p/h=22.7$ to resolve the interstitial flow, though we did not see any appreciable difference in the bulk flow properties for a coarser discretization of $D_p/h=17.0$.

We generated the initial sediment bed using a precursor simulation, in which we randomly distributed $N_{p,m}$ particles in a computational domain with periodic $x$- and $z$-boundaries above a layer of $N_{p,f}$ fixed particles, which we arranged in a hexagonal packing with random heights varying from $0 < y_0 < D_p$.  These fixed particles were used to avoid over-idealized smooth conditions at the lower wall.  We subsequently allowed the non-fixed particles to settle under ``dry" conditions, i.e. without considering hydrodynamic forces.
We then applied a large pressure gradient to produce a fluid flow rate 8 times that of the final desired flow rate, mobilizing the entire bed.  This mobilization also caused the bed to dilate, or have the average local volume fraction decrease, which in turn decreased $h_f$.  Once $h_f$ dropped to about $0.15D_p$ below the desired value, we immediately decreased the flow rate to the final flow rate reported in Table~\ref{tab:aussillous_flow_rate}, which is defined as
\begin{linenomath*}
\begin{equation}\label{eq:qf}
q_f = \frac{1}{L_x L_z}\int_0^{L_z} \int_0^{L_y} \int_0^{L_x} (1-\phi) u \, \mrm{d}x \, \mrm{d}y \, \mrm{d}z \qquad ,
\end{equation}
\end{linenomath*}
where $\phi$ is the particle volume fraction.  We adopted this procedure because we noticed a hysteresis in the particle flux between an increased flow rate and a decreased flow rate, which has also been observed by \cite{clark2015}.  Note that this procedure more closely resembles the experiments, where the particle bed is largely mobilized and then settles into a lower particle flux.

However, one problem with this procedure is that we cannot determine the final bed height \textit{a priori}.  The dilation and contraction accompanying the two flow rates is difficult to predict without executing an iterative procedure of running simulations with varying numbers of particles.  Due to the computational costs of the simulations, we did not iterate on this method, but instead accepted the values we obtained for $h_f$, which, with the exception of case A10M, are larger than those in the experiments, as seen in Table~\ref{tab:aussillous_flow_rate}.

\subsection{Comparison of wall-normal profiles}

\setlength{\unitlength}{1cm}
\begin{figure}[t]
\begin{picture}(7,8.6)
	\put(0,   4.4 ){\includegraphics[width=0.50\textwidth]{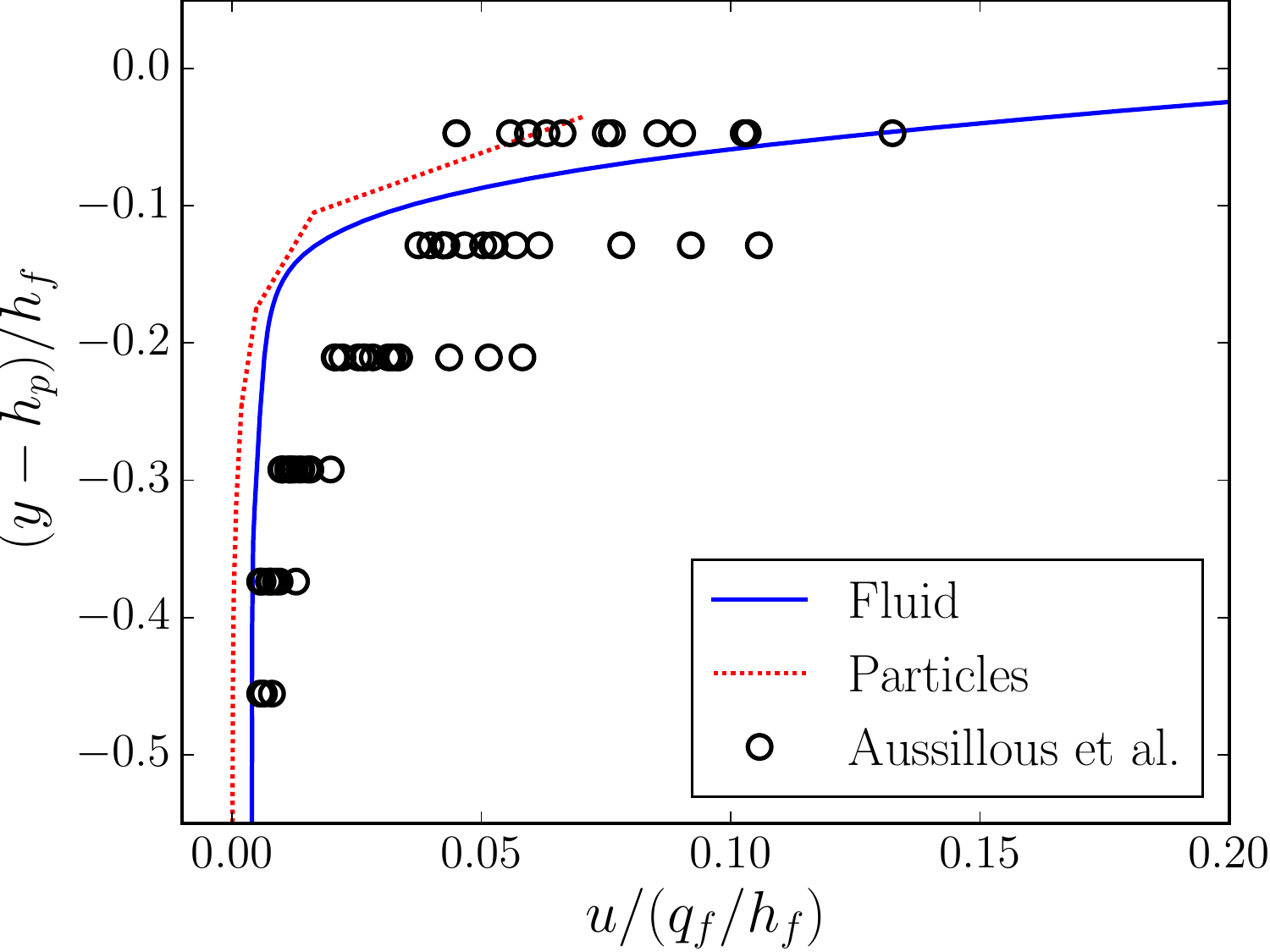}}
	\put(6.5, 4.4 ){\includegraphics[width=0.50\textwidth]{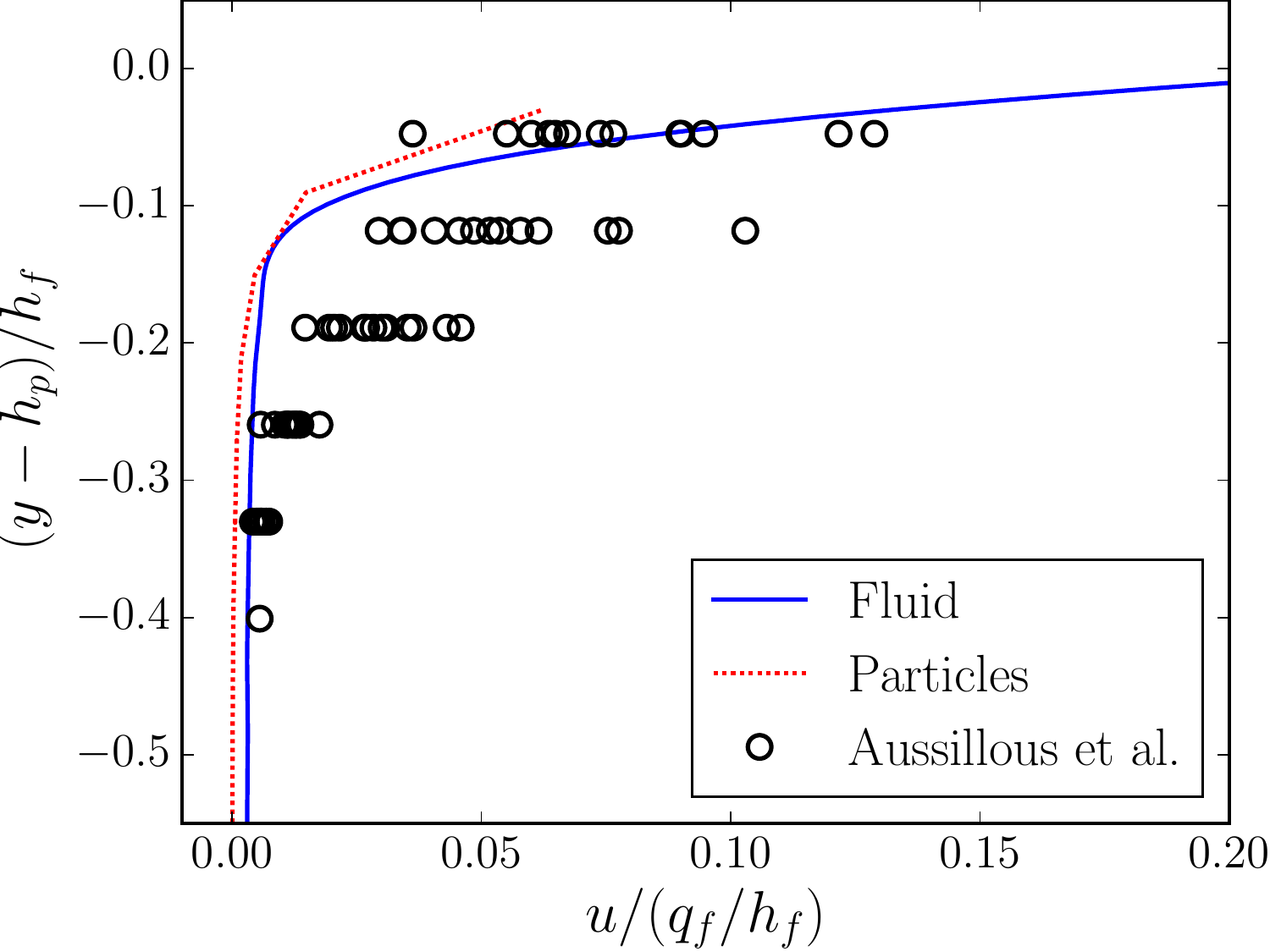}}
	\put(0,  -0.2 ){\includegraphics[width=0.50\textwidth]{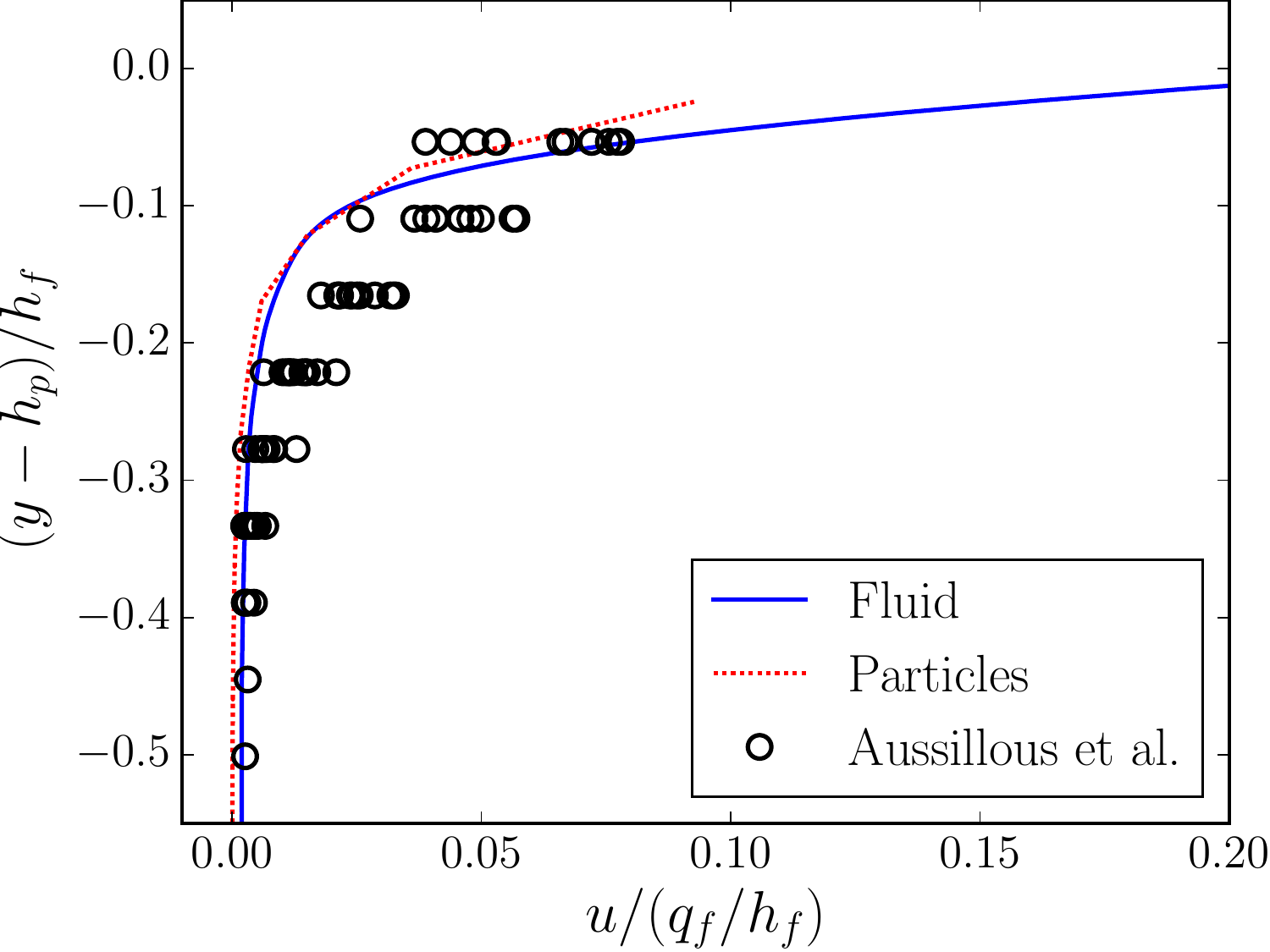}}
	\put(6.5,-0.15){\includegraphics[width=0.50\textwidth]{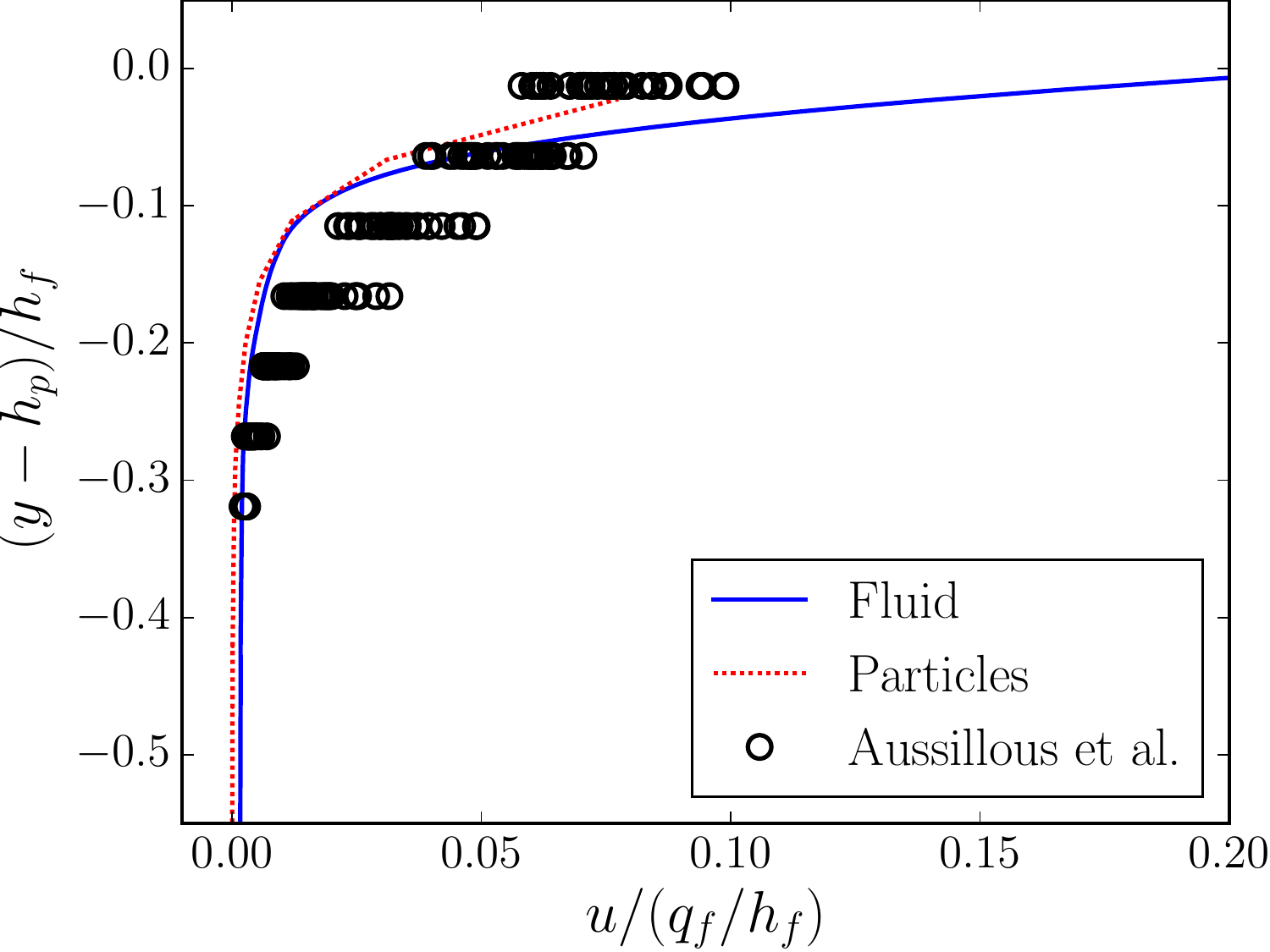}}
	\put( 0.0,  8.6)     {a) }
	\put( 6.5,  8.6)     {b) }
	\put( 0.0,  4.0)     {c) }
	\put( 6.5,  4.0)     {d) }
\end{picture}
\caption{\small \textit{Wall-normal profiles of average fluid and particle velocities near the particle/fluid interface compared to the wall-normal particle velocity profile from \cite{aussillous2013}.  a) Case A1, b) Case A2 c) Case A9 d) Case A10.}}
    \label{fig:comparison_aussillous_zoom}
\end{figure}
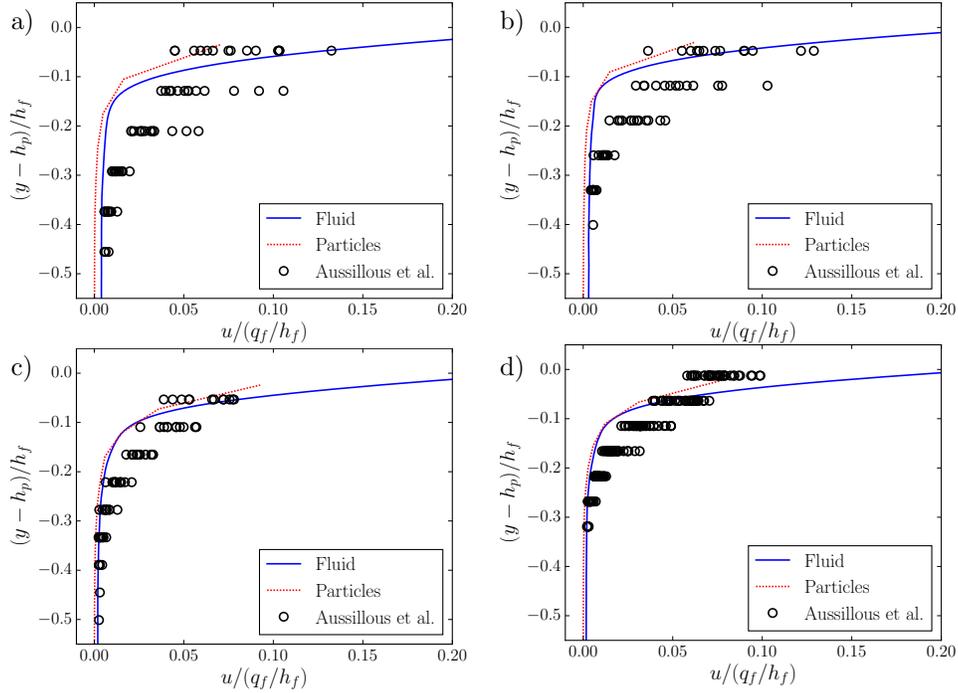

In Figure~\ref{fig:comparison_aussillous_zoom} we compare the particle velocity profiles of the simulation to the experimental results of \cite{aussillous2013}.  We calculated the particle velocity profile $u_p(y)$ from our simulations by averaging the velocities of all the particles in the streamwise and spanwise directions whose center fell within a given range of heights.  We used bins of width $R_p$ arranged such that the topmost bin extended from $y=h_p-R_p$ to $y=h_p$.
We evaluated the fluid velocity profile by averaging the $u$-velocity field in the streamwise and spanwise directions for each grid cell of the $y$-coordinate.  For this calculation, we used the particle cell volume fractions $\phi$ to exclude fluid velocities existing within the particles:
\begin{linenomath*}
\begin{equation}
\langle u \rangle_{xz} = \frac{\int_0^{L_z} \int_0^{L_x} (1-\phi) u \, \mrm{d}x \, \mrm{d}z}{\int_0^{L_z} \int_0^{L_x} (1-\phi) \, \mrm{d}x \, \mrm{d}z}
\end{equation}
\end{linenomath*}
The fluid velocity profiles exhibit a parabolic shape in the clear fluid above the bed, as shown in Figure~\ref{fig:aussillous1}.  At the interface between the clear fluid and particle bed, we observe some slip between the fluid and the particles, but within the bed the two velocity profiles are very similar, with only a slight difference due to flow between the particles.  The particle velocity profiles from the simulations compare very well with the experiments for cases A9 and A10, and reasonably well for cases A1 and A2.

Part of the discrepancy between our experiments and the simulations is due to the differences in bed heights and Shields numbers as seen in Table~\ref{tab:aussillous_flow_rate}.  In this table, we can see that cases A1 and A2 exhibit the largest differences in the fluid height between the simulations and experiments, which may have resulted in the larger deviations in the velocity profiles seen in Figure~\ref{fig:comparison_aussillous_zoom}.  Likewise, for these two cases we can also see larger differences in the Shields number, which can be sensitive to the fluid height $h_f$.

\subsection{Comparison of bulk quantities}

\begin{table}
\centering
\begin{tabular}{ccccc}
\hline
\hline
Case	& \multicolumn{3}{c}{Experimental value}		& Simulation value \\
		& $\langle q_v \rangle_T^-/q_f$	& $\langle q_v \rangle_T/q_f$		& $\langle q_v \rangle_T^+/q_f$	& $\langle q_v \rangle_T/q_f$ \\
\hline
A1		& 5.e-3			& 1.2e-2		& 1.9e-2		& 6.71e-3 \\
\hline
A2		& 5.e-3			& 1.1e-2		& 1.6e-2		& 5.17e-3 \\
\hline
A9		& 4.87e-3		& 8.20e-3		& 1.15e-2		& 7.61e-3 \\
\hline
A10		& 5.17e-3		& 7.04e-3		& 8.91e-3		& 5.84e-3 \\
\hline
A10M	& 5.17e-3		& 7.04e-3		& 8.91e-3		& 7.56e-3 \\
\hline
\hline
\end{tabular}
\caption{\small \textit{Comparison of the velocity flux $\langle q_v \rangle_T$ between our simulations and the experiments of \cite{aussillous2013}.  $\langle q_v \rangle_T^+/q_f$ and $\langle q_v \rangle_T^-/q_f$ represent the mean $\langle q_v \rangle_T/q_f$ plus and minus the standard deviation over the averaging time, respectively.}}
\label{tab:comparison_aussillous}
\end{table}

We ran the simulation until it reached a constant particle velocity flux $q_v$, defined as
\begin{linenomath*}
\begin{equation}\label{eq:qv}
q_v = \int_0^{L_y} u_p(y) \, \mrm{d}y \qquad ,
\end{equation}
\end{linenomath*}
where $u_p(y)$ is the particle velocity profile as defined in the previous section.
Unlike $q_f$, which had no variability in time, $q_v$ did vary as particles occasionally locked in place or rolled over one another.  We therefore evaluated a time-averaged value of the particle velocity flux
\begin{linenomath*}
\begin{equation}\label{eq:qv_avg}
\langle q_v \rangle_T = \frac{1}{T_\textit{avg}}\int_{t_s}^{t_f} q_v \, \mrm{d}t \qquad ,
\end{equation}
\end{linenomath*}
where $t_f$ is the time at the end of the simulation, $t_s$ is the time at which the particle flux reached steady-state, and $T_\textit{avg} = t_f - t_s$ is the time interval over which the data was averaged.  The values of $T_\textit{avg}$ are given in Table~\ref{tab:aussillous_flow_rate}.

In Table~\ref{tab:comparison_aussillous}, we can see a good agreement between our numerical results and the experimental values of the velocity flux $q_v$.  Because these quantities are derived from the particle velocity profiles, we expect to see the similar trends, namely that we underestimate the mean values from the experiments and obtain better matches for cases A9 and A10.  However, our results still fall within the margin of error of the experiments.

\subsection{Polydisperse flow}

\begin{figure}[t]
\centering
\includegraphics[width=0.65\textwidth]{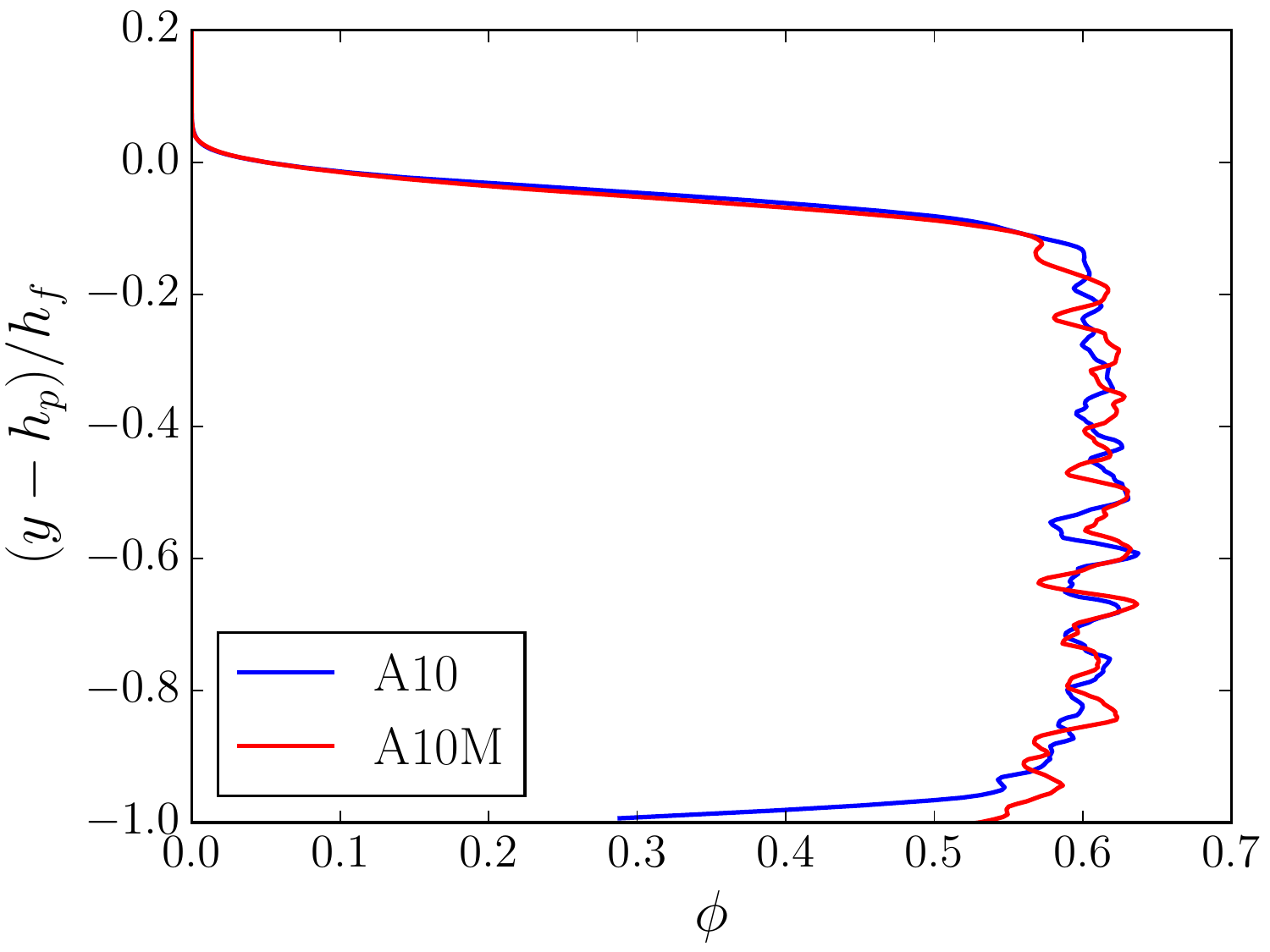}
\caption{\small \textit{Wall-normal profiles of average particle volume fractions.}}
\label{fig:aussillous_vf}
\end{figure}

\begin{figure}[t]
\centering
\includegraphics[width=0.65\textwidth]{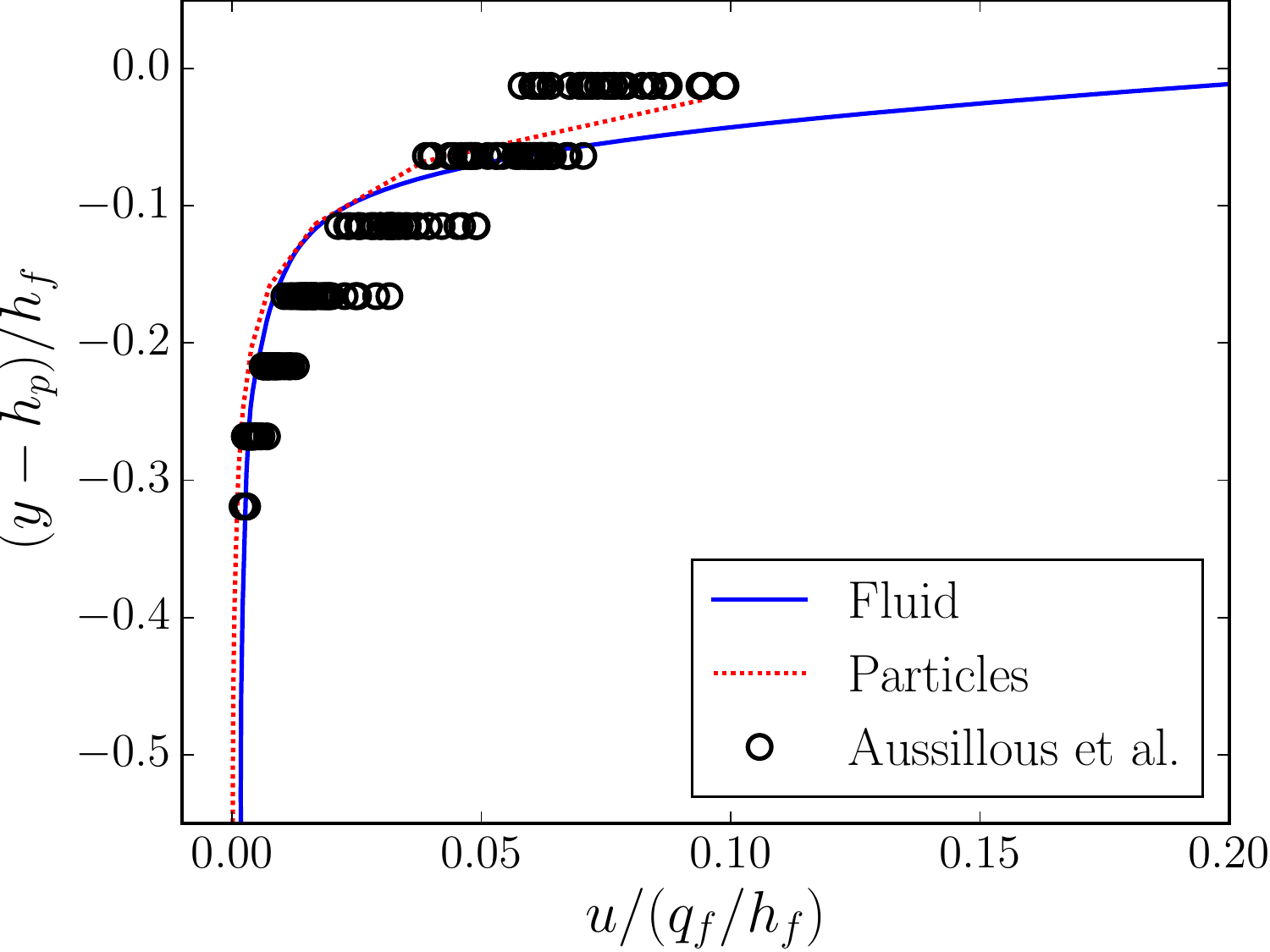}
\caption{\small \textit{Wall-normal profiles of average fluid and particle velocities for the simulation with polydisperse particles (case A10M).}}
\label{fig:aussillousM_zoom}
\end{figure}

Furthermore, we conducted another simulation to show the effect of polydispersity. In experiments, it is impossible to have a perfectly monodisperse set of particles. In their article, \cite{aussillous2013} reported having a set of spheres with diameters following a Gaussian distribution of mean $D_p = 1.1$mm and standard deviation $\sigma(D_p) = 0.1$mm, which is almost 10\% of the mean. We created a simulation containing this distribution of particle diameters and a similar submergence depth to that of case A10.  The parameters used are listed under case A10M in Table~\ref{tab:aussillous_flow_rate}.

In Figure~\ref{fig:aussillous_vf}, we do not see any appreciable changes in the particle bed volume fractions between cases A10 and A10M.  The average volume fraction within the bed is $\phi = 0.609$, which is consistent with a random sphere packing fraction.  On the other hand, in Figure~\ref{fig:aussillousM_zoom}, we see a slightly increased velocity profile compared to that of case A10 (Figure~\ref{fig:comparison_aussillous_zoom}d).  This is likely due to the decreased value of $h_f$ compared to that of A10, which results in a higher Shields number, as shown in Table~\ref{tab:aussillous_flow_rate}.  Therefore, we also obtain a velocity flux that overpredicts the mean experimental value, as shown in Table~\ref{tab:comparison_aussillous}.  However, the particle velocity profile and velocity flux still agree very well with the experimental results, and the results suggest that using monodisperse spheres is a valid approximation to polydisperse spheres for this experimental setup.

\section{Conclusions}\label{sec:conclusions}
In the present study, we presented and validated a contact model for the purpose of phase-resolved Direct Numerical Simulations, in which the disperse phase is represented by the Immersed Boundary Method. The present modeling approach allows for actual particle contact and takes all relevant contact forces into account without introducing parameters that require arbitrary calibration. These forces include lubrication forces for small inter-particle gaps, normal repulsive forces to resolve inelastic collisions, and tangential forces to represent particle friction. We demonstrated that an improved integration scheme is necessary to obtain consistent results for particle-wall collisions. Subsequently, we presented enhancements that extend the model to deal with simulations of flows over dense granular sediments. It turns out that these enhancements are crucial in order to deal with thick sediment packings. The measures taken allow us to generate sediment packings several diameters thick that are numerically stable as the packing reaches a steady-state condition. The simulations are performed by retaining the full momentum balance of a particle in enduring contact, which includes the hydrodynamic forces and the buoyant weight of a particle. Including these forces is crucial to represent phenomena like erosion and resuspension of particles. Moreover, the enhanced model allows for rolling and sliding contact, distinguishing between sticking and sliding conditions.

Altogether, the present approach yielded satisfactory agreement with the benchmark test cases for binary collisions as well as the collective motion of particles for a horizontal flow over a dense granular packing. In addition, a first test case involving polydisperse sediment was presented. The high degree of accuracy achieved is going to enable us to analyze phase-resolved numerical simulation data in great detail. Although not explicitly stated, we believe that the present approach is also applicable for the situation of vertical channel flows as well as neutrally buoyant particles in laminar and turbulent conditions. It can therefore provide a valuable tool to generate high-fidelity data even on the grain scale of any kind of multiphase flows involving rigid spheres.

\section{Acknowledgements}

This research is supported in part by the Department of Energy Office of Science Graduate Fellowship Program (DOE SCGF), made possible in part by  the American Recovery and Reinvestment Act of 2009, administered by ORISE-ORAU under contract no. DE-AC05-06OR23100.  It is also supported by the Petroleum Research Fund, administered by the American Chemical
Society, grant number 54948-ND9.  BV gratefully acknowledges the Feodor-Lynen scholarship provided by the Alexander von Humboldt foundation, Germany, and EM thanks Petrobras for partial support. The authors thank \'{E}. Guazelli and P. Aussillous for stimulating discussions on their experimental work. P. Gondret, N. Mordant, and A. Ten Cate are acknowledged for kindly providing their data.  Computational resources for this work used the Extreme Science and Engineering Discovery Environment (XSEDE), which was supported by the National Science Foundation, USA, Grant No. TG-CTS150053.

\section*{References}

\bibliography{vowinckel_bibliography}

\newpage
\begin{appendix}

\section*{Appendix}
\section{Definitions for particle-particle and particle-wall collisions}\label{app:definitions}

In order to discuss collisions in a general manner, we provide definitions for several variables that describe the contact.  Some definitions will depend on whether the interaction is between particle $p$ and a wall (particle-wall interaction, or P-W) or between particle $p$ and particle $q$ (particle-particle interaction, or P-P).  For most of the definitions, collisions between a fixed particle and a mobile particle are handled identically to particle-particle collisions, unless indicated otherwise (particle-fixed, or P-F).

\begin{itemize}

\item $R_{\textit{eff}}$ -- effective radius
	\begin{IEEEeqnarray}{rCl"s} \label{eq:R_eff} \IEEEyesnumber\IEEEyessubnumber*
		R_{\textit{eff}} &=& \frac{R_p \, R_q}{R_p + R_q} & (P-P) \\
		R_{\textit{eff}} &=& R_p & (P-W)
	\end{IEEEeqnarray}

\item $m_{\textit{eff}}$ -- effective mass
	\begin{IEEEeqnarray}{rCl"s}\label{eq:m_eff} \IEEEyesnumber\IEEEyessubnumber*
		m_{\textit{eff}} &=& \frac{m_p \, m_q}{m_p + m_q} & (P-P) \\
		m_{\textit{eff}} &=& m_p & (P-W, P-F)
	\end{IEEEeqnarray}

\item $\mbf{x}_w$ -- point on wall closest to particle

\item $\mbf{n}$ -- unit vector normal to the surface of contact, points from $\mbf{x}_p$ to $\mbf{x}_q$ (P-P) or directly towards the wall (P-W)
	\begin{IEEEeqnarray}{rCl"s} \label{eq:n} \IEEEyesnumber\IEEEyessubnumber*
		\mbf{n} &=& \frac{\mbf{x}_q - \mbf{x}_p}{||\mbf{x}_q - \mbf{x}_p||} & (P-P)\\
		\mbf{n} &=& \frac{\mbf{x}_w - \mbf{x}_p}{||\mbf{x}_w - \mbf{x}_p||} & (P-W)
	\end{IEEEeqnarray}


\item $\zeta_n$ -- distance between surfaces of the two bodies (negative value indicates overlap)
	\begin{IEEEeqnarray}{rCl"s} \label{eq:zeta_n} \IEEEyesnumber\IEEEyessubnumber*
		\zeta_n &=& ||\mbf{x}_q - \mbf{x}_p|| - R_p - R_q & (P-P)\\
		\zeta_n &=& ||\mbf{x}_w - \mbf{x}_p|| - R_p & (P-W)
	\end{IEEEeqnarray}

\item $\mbf{x}_{cp}$ -- location of contact point between surfaces, halfway between surface overlap (P-P)
	\begin{IEEEeqnarray}{rCl"s} \IEEEyesnumber\IEEEyessubnumber*
		\mbf{x}_{cp} &=& \mbf{x}_p + \left( R_p + \frac{\zeta_n}{2} \right) \mbf{n} & (P-P)\\
		\mbf{x}_{cp} &=& \mbf{x}_w & (P-W)
	\end{IEEEeqnarray}

\item $R_{p,cp}$ -- radius of particle $p$ with respect to the contact point
	\begin{IEEEeqnarray}{rCl} \label{eq:R_cp}
		R_{p,cp} &=& ||\mbf{x}_{cp} - \mbf{x}_p||
	\end{IEEEeqnarray}

\item $\mbf{g}$ -- relative velocity between particle centers of mass
	\begin{IEEEeqnarray}{rCl"s} \label{eq:g} \IEEEyesnumber\IEEEyessubnumber*
		\mbf{g} &=& \mbf{u}_p - \mbf{u}_q & (P-P)\\
		\mbf{g} &=& \mbf{u}_p & (P-W)
	\end{IEEEeqnarray}

\item $\mbf{g}_{cp}$ -- relative velocity of surface contact point
	\begin{IEEEeqnarray}{rCl"s} \label{eq:g_cp} \IEEEyesnumber\IEEEyessubnumber*
		\mbf{g}_{cp} &=& \mbf{g} + R_{p,cp} (\mbs{\omega}_p \times \mbf{n}) + R_{q,cp} (\mbs{\omega}_q \times \mbf{n}) & (P-P)\\
		\mbf{g}_{cp} &=& \mbf{g} + R_{p,cp} (\mbs{\omega}_p \times \mbf{n}) & (P-W)
	\end{IEEEeqnarray}

\item $\mbf{g}_{n,cp}$ -- component of $\mbf{g}_{cp}$ normal to surface
	\begin{IEEEeqnarray}{rCl}\label{eq:g_ncp}
		\mbf{g}_{n,cp} &=& (\mbf{g}_{cp} \cdot \mbf{n}) \mbf{n}
	\end{IEEEeqnarray}

\item $\mbf{g}_{t,cp}$ -- component of $\mbf{g}_{cp}$ tangent to surface
	\begin{IEEEeqnarray}{rCl} \label{eq:g_tcp}\label{eq:g_cp_t}
		\mbf{g}_{t,cp} &=& \mbf{g}_{cp} - \mbf{g}_{n,cp}
	\end{IEEEeqnarray}

\end{itemize}

\section{Calculating the normal contact model coefficients} \label{app:ACTM_coeff}
In order to obtain the stiffness and damping coefficients $k_n$ and $d_n$, \cite{ray2015} use nonlinear transformations and a series expansion of \eqref{eq:actm_ode} to yield the following algebraic expressions:
\begin{linenomath*}
\begin{subequations}\label{eq:actm_explicit}
\begin{equation}\label{eq:lambda_explicit}
 \lambda = \frac{1}{\alpha^2\tau_{c,0}^2} \left(-\frac{1}{2}C\eta + \sqrt{\frac{1}{4}C^2\eta^2+\alpha^2\tau_{c,0}^2\eta} \right) \qquad ,
\end{equation}
\begin{equation}\label{eq:t_explicit}
 t_* = \frac{T_c}{\tau_{c,0}}\sqrt{1-A\lambda-B\lambda^2} \qquad ,
\end{equation}
\begin{equation}\label{eq:d_n_explicit}
 d_n = \frac{2 \lambda m_\textit{eff}}{t_*} \qquad ,
\end{equation}
and
\begin{equation}\label{eq:k_n_explicit}
 k_n = \frac{m_\textit{eff}}{\sqrt{u_{in}t_*^5}} \qquad ,
\end{equation}
\end{subequations}
\end{linenomath*}
where $A=0.716, B=0.830$, $C=0.744$, $\alpha = 1.111$, and $\tau_{c,0}=3.218$ are constants.  The parameter $\eta = (\text{ln} \, e_{dry})^2$ accounts for the restitution coefficient, and we measure the impact velocity to be $u_{in} = \mbf{g}_{n,cp} \cdot \mbf{n}$ at the first occurrence of $\zeta_n \leq 0$.

\section{The tangential displacement vector} \label{app:zeta_t}

Tangential models based on spring systems require a displacement as defined by \eqref{eq:zeta_t_int}, which represents the accumulated relative motion between two surfaces  We calculate $\mbs{\zeta}_t$ in a discrete sense as follows:
\begin{IEEEeqnarray}{rCl} \label{eq:zeta_t} \IEEEyesnumber\IEEEyessubnumber*
	\widetilde{\mbs{\zeta}}_t &=& \mbs{\zeta}_t^{k-1}
		- \left( \mbs{\zeta}_t^{k-1} \cdot \mbf{n} \right) \mbf{n} \label{eq:zeta_tilde} \\
	\widehat{\mbs{\zeta}}_t &=& \frac{||\mbs{\zeta}_t^{k-1}||}{||\widetilde{\mbs{\zeta}}_t||}
		\widetilde{\mbs{\zeta}}_t \label{eq:zeta_hat} \\
	\mbs{\zeta}_t^k &=& \widehat{\mbs{\zeta}}_t + 2 \alpha_k \Delta t \, \mbf{g}_{t,cp} \qquad .
\end{IEEEeqnarray}
Equations~\eqref{eq:zeta_tilde} and~\eqref{eq:zeta_hat} rotate the displacement from the previous timestep onto a plane tangent to the two surfaces.  \cite{luding2008} implemented this rotation to account for the change in reference frame that can take place between two timesteps.  Without this rotation, the linear spring could contribute to the normal force acting between two particles.

Furthermore, when the two surfaces slip according to the Coulomb friction criteria, the displacement vector should not grow as the two surfaces continue to slide past one another.  Instead, we reset the displacement to that which achieves the Coulomb friction force:
\begin{linenomath*}
\begin{equation}\label{eq:zeta_t_coulomb}
\mbs{\zeta}_t = -\frac{ ||\mu \mbf{F}_n|| \mbf{t} + d_t \mbf{g}_{t,cp} }{ k_t }
	\qquad \mrm{if} \quad ||\mbf{F}_{t,LS}|| > ||\mu \mbf{F}_n||  \qquad .
\end{equation}
\end{linenomath*}

\end{appendix}
\end{document}